%% file: ERR_couplestress_IJES_rev_arxiv2.tex
\documentclass[a4paper,11pt]{article}

\usepackage[english]{babel}
\usepackage[latin1]{inputenc}
\usepackage[T1]{fontenc}
\usepackage{ae,aecompl} 
\usepackage{graphicx,color,setspace,amsmath}
\usepackage{amssymb}
\usepackage{natbib}
\usepackage{setspace}
\usepackage{mathrsfs}
\usepackage{lscape}
\usepackage{rotating}
\usepackage{verbatim}
\usepackage{amsfonts}  
\usepackage{amsbsy}    
\input{macro}

\begin{document}
\centerline{\Large \textbf{On fracture criteria for dynamic crack propagation}}

\centerline{\Large  \textbf{in elastic materials with couple stresses} }

\begin{center}
L. Morini$^{(1),(2),*}$, A. Piccolroaz$^{(2)}$, G. Mishuris$^{(3)}$ and E. Radi$^{(1)}$\\
\end{center}

\centerline{$^{(1)}$\emph{Dipartimento di Scienze e Metodi dell'Ingegneria, Universit\'a di Modena e Reggio Emilia,}}

\centerline{\emph{Via Amendola 2, 42100, Reggio Emilia, Italy.}}

\centerline{$^{(2)}$\emph{Department of Civil, Environmental and Mechanical Engineering, University of Trento,}}

\centerline{\emph{Via Mesiano 77, 38123, Trento, Italy.}}

\centerline{$^{(3)}$\emph{Institute of Mathematical and Physical Sciences, Aberystwyth University,}}

\centerline{\emph{ Ceredigion SY23 3BZ, Wales, U.K.}}

\vspace{2cm}

\begin{abstract}
The focus of the article is on fracture criteria for dynamic crack propagation in elastic materials with microstructures. Steady-state propagation of a Mode III
semi-infinite crack subject to loading applied on the crack surfaces is considered. The micropolar behavior of the material is described by the theory of couple-stress
elasticity developed by Koiter. This constitutive model includes the characteristic lengths in bending and torsion, and thus it is able to account for the underlying
microstructures of the material. Both translational and micro-rotational inertial terms are included in the balance equations, and the behavior of the
solution near to the crack tip is investigated by means of an asymptotic analysis. The asymptotic fields are used to evaluate the dynamic
J-integral for a couple-stress material, and the energy release rate is derived by the corresponding conservation law. The propagation stability is studied according to the
energy-based Griffith criterion and the obtained results are compared to those derived by the application of the maximum total shear stress criterion.  \\

\end{abstract}

\vspace{5cm}
$^*$ Corresponding author. Tel.: +390461282583, email address: lorenzo.morini@unitn.it.

\newpage

\section{Introduction}
In many experimental analyses 
it has been shown that the mechanical behavior of brittle materials such as ceramics,
composites, cellular materials, foams, masonry, bones tissues, glassy and semicrystalline polymers, is strongly affected by the the microstructure.

The influence of materials inhomogeneities and defects on the mechanical properties of the materials and their interactions with cracks have been extensively studied in the framework of 
classical theory of elasticity by means of homogenization theories \citep{Hash1, Esh1, Bud1, BudOcon1}, and the effective elastic moduli of bodies containing 
several ensembles of microstructures have been determined using for instance self-consistent methods \citep{Kach1, Kach2, HuangHu1}. Furthermore, modern multi-scale simulation approaches have been developed for
modelling microstructural properties of the materials \citep{ AskBen1, AskGit1, Silb1, AndAv1}.

In \citet{BigDru1} it has been shown that 
classical homogenization results describe accurately elastic properties of heterogeneous materials in situations where the bodies are subjected to slowly-varying loading and the displacement and stress gradients 
are small. If high gradients are present, standard homogenized materials cannot represent the physical response of composite elastic media. Moreover, approaches based on classical elasticity 
theory cannot always predict enough accurately the size effect experimentally observed when the representative scale of the deformation field becomes comparable to the length scale of the microstructure \citep{Lakes1}. 
For example, it has been detected that in presence of stress concentration, such as near inclusions and holes, the strength of the material is higher if the grain size is smaller, and that the bending and torsional 
strength of beams and wires are greater if their cross-section is thinner \citep{FleckMul1}. Since stress concentration factors derived applying standard homogenization procedures to Cauchy elastic materials depend 
only on the shape of the inhomogeneities and not on its size, they cannot describe accurately these effects. On the other hand, the utilization of multi-scale techniques for studying microstructural properties 
of the materials implies challenging numerical computations, and the validation of the results requires a critical comparison with analytical solutions and experimental data.

 Generalized theories of continuum mechanics, such as micropolar elasticity \citep{Coss1}, indeterminate couple stress elasticity \citep{Koit1} and strain gradient theories \citep{FleckHutch1, Aif1, DalWil1},
may be considerate as an effort to include rigorously defined characteristic length scales and to study influence of microstructures on the material behavior 
avoiding strong numerical calculations required by multi-scale approaches. Indeed, exact analytical formulas for characteristic lengths in couple stress elastic materials have been derived 
via higher order homogenization of heterogeneous Cauchy elastic materials by \citet{BigDru1} and via numerical computations by \citet{AskAif1}. 
Analytical solutions derived for couple stress and strain gradient solids can also be used for validating results of numerical simulations and analyzing experimental data obtained for materials with microstructures \citep{Lakes2}.

Indeterminate couple stress elasticity theory developed by \cite{Koit1} provides two distinct characteristic length scales for bending and torsion. Moreover, it includes the effects of the microrotational 
inertia, which can be considered as an additional dynamic length scale. Therefore, in order to study crack propagation stability in couple-stress elastic materials, new fracture 
criteria accounting for both effects of scale lengths and microrotational inertia must be formulated \citep{Mor1}.
For antiplane crack problems, in \citet{Geo1} and \citet{Radi1} a critical level $\tau_{C}$ for the maximum total shear stress ahead of the crack tip at which the crack starts propagating 
has been proposed as fracture criterion. This is known as the maximum total shear stress criterion, and later in the article we will refer to it as the $t^{\textrm{max}}$ criterion. 

The J-integral for static crack problems in couple stress elasticity has been derived by \citet{LubMark1} in the case of an homogeneous material, and by \citet{PiccMish5} for an interfacial 
crack. The energy release rate can be evaluated by means of the conservation of this integral, and further in the paper we will refer to it as ERR. Nevertheless, energy-based dynamic fracture criteria for this
kind of materials are still unknown in literature. For that reason, the principal aim of this paper is to generalize the static energy release rate expression to the case of dynamic steady state crack
propagation, and to study the effects of the microstructure on the propagation stability by applying the energy Griffith criterion \citep{Freund1}. The results are compared with those obtained
in \citet{MishPicc1}, where the $t^{\textrm{max}}$ criterion has been adopted. 

The structure of the paper is organized as follows: in Section \ref{ss_crack} the problem of a semi-infinite Mode III steady state propagating
crack in couple stress elastic materials is formulated. Both translational and micro-rotational inertial terms are included in the balance equations, and a distributed loading applied on the crack surfaces is assumed. 
In Section \ref{dynamic_ss}, the dynamic conservation laws derived by \cite{FreuHutch1}  and \cite{Freund1} for classical elasticity are generalized to couple stress materials. 
General expressions for the dynamic J-integral and energy release rate associated to steadily propagating cracks in couple stress elastic solids are derived. Explicit forms are obtained 
for the case of a Mode III crack, and the path-independence of the J-integral is demonstrated in Appendix A.

An asymptotic analysis of the stress and displacement fields near to the crack tip is performed in Section \ref{Asym}. The contribution of the asymptotic terms to the dynamic
energy release rate is analyzed in details, the leading term corresponding to finite non-zero energy is individuated and an explicit formula for the J-integral evaluated along a 
circular path surrounding the crack tip is derived. The obtained formula involves a constant term depending on the boundary conditions of the problem and indicating the 
amplitude of the leading order term of the asymptotic shear stress. This term is evaluated in closed form in Section \ref{J_ev} by performing an asymptotic expansion of the full-field solution derived
in \citet{MishPicc1} for the same loading configuration applied at crack faces. In this Section, the asymptotics expansion of the full-field solution is also used for deriving an alternative equivalent formula for 
the dynamic J-integral, calculated considering the square-shaped path around the crack tip introduced by \citet{Freund1} and used in \citet{Geo1, GourGeo1, AraGian1}. The energy 
release rate associated to a steady propagating Mode III crack in couple stress elastic solids is compared to the corresponding expression in classical elastic materials.

In Section \ref{results}, the obtained expression for the energy release rate is used for studying subsonic crack propagation stability. Assuming the energy-based Griffith criterion 
\citep{Willis1, Willis2, ObrMov1}, under the considered loading conditions the steady state propagation turns out to be unstable regardless 
of the values the microstructural parameters. This result appears to be in contrast with those detected in \cite{MishPicc1} adopting the  $t^{\textrm{max}}$ criterion, which instead 
shows a stabilizing effect in presence of relevant microstructures contribution. In the authors' opinion, this discrepancy may be due to the fact that the energy release rate  
depends only on the leading term of the asymptotic expansion of the stresses, which dominates very close to the crack tip but provides unphysical features such as negative total shear stress ahead of 
the crack tip. Therefore, at a characteristic distance from the crack tip the sole leading order term may not describe the correct behavior of stresses and displacements. Differently, the total shear stress involved 
in the $t^{\textrm{max}}$ criterion is calculated by means of the full-field solution, that takes fully into account the microstructural contributions. As a consequence, in order to study the crack propagation
stability in elastic materials with microstructure, fracture criteria including also higher order terms of the asymptotic stresses and involving two or more characteristic parameters should be used. 
Note that in classical elasticity fracture criteria including higher order terms contributions such as T-stress criterion \citep{HancDu1,SmithAya1} have also been proposed.

\section{Steady-state cracks in couple stress elastic materials}
\label{ss_crack}
In this Section the problem of the steady-state dynamic propagation of a Mode III crack in elastic materials with microstructures is formulated by means of the fully
dynamical version of the couple-stress elastic model, accounting both translational and micro-rotational inertial terms into the balance equations.
Reference is made to a fixed Cartesian coordinate system $(0, x_1, x_2, x_3)$ centred at the crack tip at the initial time $t=0$.
Under antiplane shear deformation, the indeterminate theory of couple stress elasticity \citep{Koit1} adopted in the present study provides the following kinematical
compatibility conditions between the out-of-plane displacement $u_3$, rotation vector $\BGvf$, strain tensor $\BGve$ and rotation gradient tensor $\BGc$:
\beq
\Gve_{13}=\fr{1}{2}u_{3,1},\quad \Gve_{23}=\fr{1}{2}u_{3,2},\quad \Gvf_{1}=\fr{1}{2}u_{3,2},\quad \Gvf_{2}=-\fr{1}{2} u_{3,1},
\eequ{eps_phi}
\beq
\Gc_{11}=-\Gc_{22}=\fr{1}{2} u_{3,12},\quad \Gc_{21}=-\fr{1}{2}u_{3,11},\quad \Gc_{12}=\fr{1}{2}u_{3,22}.
\eequ{chi}

 Therefore, the rotations are derived from displacement. The rotation gradient tensor $\BGc$, also known as deformations curvature tensor or torsion-flexure tensor \citep{Koit1}, 
is defined in the general three-dimensional case as $\BGc=\nabla\BGvf=\nabla\times\BGve$. The vanishing of the Saint Venant tensor (or incompatibility tensor) requires $\BGc$ to be irrotational:
\beq
\nabla\times\BGc=\nabla\times\nabla\times\BGve=0.
\eequ{irrot}
Using expressions \eq{chi}, it can be immediately verified that relation \eq{irrot} is satisfied. According to the indeterminate couple stress theory the non-symmetric Cauchy stress tensor $\Bt$ can 
be decomposed into a symmetric part $\BGs$ and a skew-symmetric part $\BGj$, namely $\Bt=\BGs+\BGj$. In addition, the couple stress tensor $\BGm$ is introduced as the work-conjugated quantity of $\BGc^T$. 
The reduced tractions vector $\Bp$ and the couple stress tractions vector $\Bq$ are defined as
\beq
\Bp=\Bt^T\Bn+\fr{1}{2}\nabla\Gm_{nn}\times\Bn, \qquad \Bq=\BGm^T\Bn-\Gm_{nn}\Bn,
\eequ{trac}
respectively, where $\Bn$ denotes the outward unit normal and $\Gm_{nn}=\Bn\cdot\BGm\Bn$. The conditions of dynamic equilibrium of forces and moments, taking into consideration rotational inertia, and
neglecting body forces and body couples, write
\beq
\Gs_{13,1}+\Gs_{23,2}+\Gj_{13,1}+\Gj_{23,2}=\Gr\ddot{u}_3,\quad \Gm_{11,1}+\Gs_{21,2}+2\Gj_{23}=J\ddot{\Gvf}_1,\quad \Gm_{12,1}+\Gs_{22,2}-2\Gj_{13}=J\ddot{\Gvf}_2,
\eequ{balance_eq}
where $\BGr$ is the mass density and $J$ is the rotational inertia.

Within the context of small deformations theory, the total strain $\BGve$ and the deformation curvature $\BGc$ are connected to stress and couple stress through the following isotropic constitutive relations
\beq
\BGs=2G\BGve+\lambda (\mbox{tr}\BGve)\BI, \qquad \BGm=2G\ell^2(\BGc^T+\Gn\BGc),
\eequ{const_rel}
where $G$ is the elastic shear modulus, $\ell$ and $\eta$ the couple stress parameters, with $-1<\eta<1$. Note that for antiplane deformations $\mbox{tr}\BGve=0$.
Both material parameters $\ell$ and $\Gn$ depend on the microstructure and can be connected to the material characteristic length in bending and in torsion, namely
\beq
\ell_b=\ell/\sqrt{2},\qquad \ell_t=\ell\sqrt{1+\Gn}.
\eequ{char_l}

Typical experimental values of $\ell_b$ and $\ell_t$ for some classes of materials with microstructure can be found in \citet{Lakes1,Lakes2}, and analytical expressions
for these moduli have been derived via homogenization of heterogeneous Cauchy elastic materials by \citet{BigDru1}.The limit value of $\Gn=1$ corresponds to vanishing characteristic length in
torsion, which is typical of polycristalline metals. Moreover, from the definitions \eq{char_l} it follows that $\ell_t=\ell_b$ for $\Gn=0.5$ and $\ell_t=\ell_b=\sqrt{2}$ for $\Gn=-1$. The constitutive
equations of the indeterminate couple stress theory do not define the skew-symmetric part $\BGj$ of the total stress tensor $\Bt$, which instead is determined by the equilibrium equations
\eq{balance_eq}$_{2,3}$. Constitutive equations \eq{const_rel} together with the compatibility relations \eq{eps_phi} and \eq{chi} give stresses and couple stresses in terms of the out of plane
displacement $u_3$:
\beq
\Gs_{13}=Gu_{3,1},\qquad \Gs_{23}=Gu_{3,2},
\eequ{sigma}
\beq
\Gm_{11}=-\Gm_{22}=G\ell^2(1+\Gn)u_{3,12}, \quad \Gm_{21}=G\ell^2(u_{3,22}-\Gn u_{3,11}), \quad \Gm_{12}=-G\ell^2(u_{3,11}-\Gn u_{3,22}).
\eequ{mu}

The introduction of \eq{mu} into \eq{balance_eq}$_{2,3}$ yields:
\beq
\Gj_{13}=-\fr{G\ell^2}{2}\GD u_{3,1}+\fr{J}{4}\ddot{u}_{3,1}, \qquad \Gj_{23}=-\fr{G\ell^2}{2}\GD u_{3,2}+\fr{J}{4}\ddot{u}_{3,2},
\eequ{tau}
where $\GD$ denotes the Laplace operator. By means of \eq{sigma} and \eq{tau}, the equation of motion \eq{balance_eq}$_1$ becomes
\beq
\GD u_3-\fr{\ell^2}{2}\GD\GD u_3= \fr{\Gr}{G}\ddot{u}_{3}-\fr{J}{4G}\GD\ddot{u}_{3}.
\eequ{eq_motion}

We assume that the crack propagates with a constant velocity $v$ along the $x_1$-axis. In this case it is convenient to introduce a moving framework $x=x_1-vt, y=x_2, z=x_3$, and the out of the plane
displacement can be assumed in the form:
\beq
u_{3}(x_1,x_2,t)=w(x,y).
\eequ{w}

It follows that the time derivative of the displacement $w$ can be written in terms of the derivative with respect to $x$, namely $\dot{w}=-vw_{,x}$ and thus $\ddot{w}=v^2w_{,xx}$. Therefore the equation
of motion \eq{eq_motion} under steady-state conditions becomes:
\beq
\GD w-\fr{\ell^2}{2}\GD\GD w= m^2\left({w}_{,xx}-h_0^2\ell^2\GD w_{,xx}\right)
\eequ{eq_steady}
where $m=v/c_s$ is the normalized crack velocity, $c_{s}=\sqrt{G/\Gr}$ is the shear wave speed for classical elastic materials, the characteristic length $h$ is defined as $h=c_s/\Gf$ with $\Gf=\sqrt{4G/J}$,
and $h_0=h/\ell$ (see \citet{MishPicc1}).

According to \eq{trac}, the non-vanishing components of the reduced traction and couple stress traction vectors along the crack line $y=0$ can be written as
\beq
p_3=t_{23}+\fr{1}{2}\Gm_{22,x}, \qquad q_1=\Gm_{21},
\eequ{tract0}
respectively. By using \eq{sigma}$_{2}$, \eq{mu}$_{1,2}$, \eq{tau}$_{2}$ and \eq{tract0}, the skew-symmetry of the Mode III crack problem requires ahead of the crack tip:
\beq
w=0, \quad w_{,yy}-\Gn w_{,xx}=0, \quad \mbox{for} \quad x>0,\;y=0.
\eequ{bound+}

On the crack surface, vanishing of the reduced traction and couple stress traction yield to the following boundary conditions for the function $w$:
\beq
w_{,y}-\fr{\ell^2}{2}\left[\left(2+\Gn-2m^2h_0^2\right)w_{,xx}+w_{,yy}\right]_{,y}=-\fr{1}{G}\tau(x)\quad w_{,yy}-\Gn w_{,xx}=0, \quad \mbox{for} \quad x<0,\;y=0,
\eequ{bound-}
where $\tau(x)$ is the loading applied on the crack faces, which is assumed to have the following form:
\beq
\tau(x)=\fr{T_0}{L}e^{x/L}, \quad x<0.
\eequ{loading}
Note that although we discuss here only a specific loading condition, the main conclusions reported in this paper have been confirmed for other types of loading.

\section{Dynamic energy release rate}
\label{dynamic_ss}
In this Section, the dynamic conservation laws obtained for linear elastic media by \cite{Freund1} and  \cite{FreuHutch1} are generalized to couple stress elastic materials.
An explicit integral expression for the dynamic energy release rate associated to steady state cracks propagation in elastic solids with presence of couple stress is derived.

The energy release rate for a dynamic crack has been defined by \cite{Freund1} as the following limit:
\beq
\CE=\lim_{\Gamma\rightarrow 0}\left[\fr{F(\Gamma)}{v}\right],
\eequ{ERR_def}
where $F$ is the total energy flux through the contour $\Gamma$ surrounding the crack tip. For couple stress elastic materials, the energy flux for a dynamic crack propagating along $x_1$-axis is given by:
\beq
F(\Gamma)=\int_{\Gamma}\left[(W+T)vn_1+\Bt^T\Bn\cdot \fr{\partial\Bu}{\partial t}+\BGm^T\Bn\cdot\fr{\partial \BGvf}{\partial t}\right]ds,
\eequ{CS_flux}
where $\Bn$ is an outward unit normal on $\Gamma$, $W$ denotes the strain-energy density
\beq
W=\fr{1}{2}\left(\BGs\cdot\nabla\Bu+\BGm^T\cdot\nabla\BGvf\right)=G\BGe\cdot\BGe+G\ell^2(\BGc\cdot\BGc+\Gn\BGc\cdot\BGc^T),
\eequ{strain_energy}
and $T$ is the kinetic energy density
\beq
T=\fr{1}{2}\left(\rho\left|\dot{\Bu}\right|^2+J\left|\dot{\BGvf}\right|^2\right).
\eequ{kin_energy}

Since a Mode III steady-state crack propagating at constant velocity $v$ along the $x_1$-axis is considered, the expression \eq{w} for the out-of-plane displacement is used and then in the moving
framework $(x,y,z)$ the energy flux \eq{CS_flux} assumes the special form
\beq
F(\Gamma)=v\int_{\Gamma}\left[(W+T)n_x-\Bt^T\Bn\cdot\Be_z w_{,x}-\BGm^T\Bn\cdot\BGvf_{,x}\right]ds,
\eequ{CS_flux1}
then the generalized J-integral for an antiplane dynamic steady-state crack in couple-stress elastic materials can be defined:
\begin{eqnarray}
\CJ=\fr{F(\Gamma)}{v} & = & \int_{\Gamma}\left[(W+T)n_x-\Bt^T\Bn\cdot\Be_z w_{,x}-\BGm^T\Bn\cdot\BGvf_{,x}\right]ds =  \label{J_intA}\\
                      & = & \int_{\Gamma}\left[(W+T)\Bn-\Bp \cdot\Be_z \nabla w-(\nabla \BGvf)^T \Bq\right]\cdot \Be_x ds   \label{J_intB}
\end{eqnarray}

The expressions \eq{J_intA} and has been proved to be path independent, the details of the demonstration are reported in Appendix A. Moreover, the equivalence of the two alternative forms of the
J-integral \eq{J_intA} and \eq{J_intB}, the first written in function of the tractions and the second of the reduced tractions, is demonstrated in Appendix B. The  \eq{J_intA}  is the generalization of the
static expressions derived by \cite{AtkLepp1,AtkLepp2}, and \cite{LubMark1} to the antiplane dynamic steady state case. The dynamic energy release rate is then defined by the limit:
 \beq
 \CE= \lim_{\Gamma\rightarrow 0}\int_{\Gamma}\left[(W+T)n_x-\Bt^T\Bn\cdot\Be_z w_{,x}-\BGm^T\Bn\cdot\BGvf_{,x}\right]ds.
 \eequ{ERR_dyn}

 Using definitions \eq{eps_phi} and \eq{chi}, the strain energy density \eq{strain_energy} becomes
\beq
W=\fr{G}{2}(w_{,x}^2+w_{,y}^2)+\fr{G\ell^2}{4}\left[(w_{,xx}+w_{,yy})^2+2(1+\Gn)(w^2_{,xy}-w_{,xx}w_{,yy})\right],
\eequ{strain_enmode3}
whereas for steady state propagation the kinetic energy density is given by
\beq
T=\fr{v^2}{2}\left[\rho w_{,x}^2+\fr{J}{4}(w_{,xy}^2+w_{,xx}^2)\right].
\eequ{ki_enmode3}
 A polar coordinates system $(r,\Gt)$ centered at the crack tip is assumed, and a circular contour of radius $r$ around the crack tip with $\Bn=(\cos\Gt,\sin\Gt,0)$ is considered. Then,
 substituting expressions \eq{strain_enmode3} and \eq{ki_enmode3} the J-integral \eq{J_intA} becomes
\begin{eqnarray}
\CJ & = & \int^{\pi}_{-\pi}\left\{\fr{v^2}{2}\left[\rho w_{,x}^2\cos\Gt+\fr{J}{4}(w_{,xy}^2+w_{,xx}^2)\cos\Gt-\fr{J}{2}w_{,x}\left(w_{,xxx}\cos\Gt+w_{,yxx}\sin\Gt\right)\right]\right.+\nonumber\\
    & + & \fr{G}{2}\left[(w_{,y}^2-w_{,x}^2)\cos\Gt-2w_{,x}w_{,y}\sin\Gt\right]+\label{Jex}\\
    & + & \left.\fr{G\ell^2}{2}\left[\fr{(\Delta w)^2}{2}\cos\Gt-\Delta w\left(w_{,xx}\cos\Gt+w_{,xy}\sin\Gt\right)+w_{,x}\left(\Delta w_{,x}\cos\Gt+\Delta w_{,y}\sin\Gt\right)\right]\right\}r d\Gt \nonumber.
\end{eqnarray}

According to the definition (\ref{ERR_dyn}), the energy release rate can be evaluated as the limit for $r\rightarrow 0$ of the integral (\ref{Jex}). In order to evaluate explicitly the J-integral (\ref{Jex})
and to investigate the variation of the energy release rate (\ref{ERR_dyn}) in function of the crack propagation velocity and its implications on
propagation stability, the behavior of the out-of-plane displacement $w$ near to the crack tip is studied by means of an asymptotic analysis in the next Section.
 
\section{Asymptotic crack tip fields}
\label{Asym}

The following standard asymptotic expression for out-of-plane displacement $w$ in separate variables form is considered:
\beq
w(r,\Gt)=r^sF_s(\Gt),\quad  r \rightarrow 0,
\eequ{asym}

We are interested in finding the terms of the asymptotic solution \eq{asym} corresponding to finite and non-zero contributions to the J-integral (\ref{Jex}) in the limit  $r \rightarrow 0$. Since the
displacement $w$ should be bounded and symmetrical, it follows immediately that $s>0$, and then no zero-order terms are present in the asymptotic expansion. Substituting the expression \eq{asym} in the generalized
J-integral formula (\ref{Jex}) and using the following derivative rules which hold for an arbitrary function $f(x,y)=f(r,\Gt)$:
\beq
f_{,x}=f_{,r}\cos\Gt-f_{,\Gt}\fr{\sin\Gt}{r}, \quad f_{,y}=f_{,r}\sin\Gt+f_{,\Gt}\fr{\cos\Gt}{r},
\eequ{deriv}
we get:
\begin{eqnarray}
\CJ & = & \fr{r^{2s-1}}{2}\int^{\pi}_{-\pi}\left\{\rho v^2(sF_s\cos\Gt-F_s^{'}\sin\Gt)^2\cos\Gt+G\left[(1+s)F_sF_s^{'}\sin\Gt-(sF_s^2+F_s^{'2})\cos\Gt\right]\right\}d\Gt+\nonumber\\
    & + & \fr{r^{2s-3}}{4}\int^{\pi}_{-\pi}\left\{G\ell^2(s^2F_s+F_s^{''})\left[(F_s-s(2-s)F_s^{''})\cos\Gt+2F_s^{'}\sin\Gt\right]\right.+\label{Jsing}\\
    & + & \left.\fr{J v^2}{2}(2-s)(sF_s\cos\Gt-F_s^{'}\sin\Gt)\left[s(s-(2-s)\cos 2\Gt)F_s+2(1-s)F_s^{'}\sin 2\Gt+2 F_s^{''}\sin^{2}\Gt \right]\right\} d\Gt \nonumber,
\end{eqnarray}
where the superscript $^{'}$ denotes the total derivative respect to the variable $\theta$.

Observing expression \eq{Jsing}, in agreement with the results reported in \cite{Radi1} for the static case, we deduce that the finiteness of the energy release rate towards
the crack tip requires that $s\geq 3/2$ for any non integer number.

Including higher order terms in the asymptotic expansion in the form $\sum_{i}r^{s_{i}}F_{s_{i}}(\Gt)$ and using this expression in (\ref{Jex}), terms of order
$r^{s_i+s_j-1}$ and $r^{s_i+s_j-3}$ where $i \:\neq j$ are detected. These terms involve both $F_{s_{i}}$ and $F_{s_{j}}$ and can have a non vanishing impact to the value of the energy release rate.
Therefore we need to consider several asymptotic terms in the form of (\ref{asym}) and analyze the possible correlation between them in the nonlinear functional (\ref{Jex}).
According with this discussion and in order to find the terms corresponding to finite and non-zero contributions to the energy release rate, we assume $1\le s<3$.

It can be easily demonstrated that if $|s_i-s_j|<2,\:\forall\: i \:\neq j$, only the leading order term of the
governing equation \eq{eq_steady} can be considered, while if more terms are required with exponents differing by 2 or more than 2, the full equation \eq{eq_steady} must be considered 
in the analysis \citep{PiccMish5}. As a consequence, assuming  $1\le s<3$, we can then keep only the leading term of the evolution equation \eq{eq_steady}
\beq
\GD(\GD w-\Gl^2w_{,xx})=0 \qquad \mbox{where} \qquad \Gl^2=2m^2h_0^2=\fr{Jv^2}{2G\ell^2}.
\eequ{eq_lead}

Introducing the expression \eq{asym} in \eq{eq_lead}, the general asymptotic solution of the equation of motion has been obtained, the derivation is illustrated in details in the Section containing the supplementary material.
Referring to this general solution, since we are assuming values in the range $1\le s<3$, the terms corresponding to $s= 1, 3/2, 2, 5/2$  needs to be considered and the asymptotic expression for 
the out of plane displacement $w$ turns out to be:  
\begin{eqnarray}
w(r,\Gt) & = & B_1r\sin\Gt+ B_2r^{3/2}\left[\sin\fr{3}{2}\Gt-(1-\Gl^2\sin^2\Gt)^{3/4}\fr{1+\Gn}{1+\Gn-\Gl^2}\sin\fr{3}{2}\GF(\Gt)\right]+\label{lead_w}\\
         & + & B_3r^2\sin2\Gt+ B_4r^{5/2}\left[\sin\fr{3}{2}\Gt-(1-\Gl^2\sin^2\Gt)^{5/4}\fr{1+\Gn}{1+\Gn-\Gl^2}\sin\fr{5}{2}\GF(\Gt)\right]+O(r^{5/2}),\nonumber 
\end{eqnarray}
where
\beq
\GF(\Gt)=\arcsin\left(\fr{\sqrt{1-\Gl^2}\sin\Gt}{\sqrt{1-\Gl^2\sin^2\Gt}}\right),
\eequ{Phi}
and the constants $B_1, B_2, B_3$ and $B_4$ define the amplitude of each asymptotic term the sum and thus must to be evaluated according to the boundary conditions 
\eq{bound+} and \eq{bound-}.

It has been verified by symbolic calculations that the terms of the \eq{lead_w} corresponding to $s=1$ and $s=2$, similarly to what has been detected by \cite{Radi1} and
\cite{PiccMish5} for a stationary crack case do not contribute to the J-integral and to energy release rate. Although these terms are relevant for evaluating displacement and total 
shear stress at the crack tip, the only finite and non-vanishing contribution to the generalized J-integral \eq{Jex} is associated to the order $s=3/2$ of the asymptotic expression
\eq{lead_w}, and then the energy release rate is given by:
\begin{eqnarray}
\CE       & = &  \lim_{r\rightarrow 0}\CJ = \fr{B_2^2 G\ell^2}{64}\int^{\pi}_{-\pi}\left\{\left(9 H_{3/2}+4 H_{3/2}^{''}\right)\left[\left(4 H_{3/2}-3 H_{3/2}^{''}\right)\cos\Gt+8H_{3/2}^{'}\sin\Gt\right]\right.+ \label{J_32}\\
          & + & \left.\lambda^2 \left(3H_{3/2}\cos\Gt-2H_{3/2}^{'}\sin\Gt\right)\left[3(1-3\cos 2\Gt)H_{3/2}-4H_{3/2}^{'}\sin 2\Gt+8 H_{3/2}^{''}\sin^{2}\Gt \right]\right\} d\Gt \nonumber,
\end{eqnarray}
where 
\begin{eqnarray}
H_{3/2}(\Gt) & = & \sin\fr{3}{2}\Gt-(1-\Gl^2\sin^2\Gt)^{3/4}\fr{1+\Gn}{1+\Gn-\Gl^2}\sin\fr{3}{2}\GF(\Gt) = \label{H32}\\
             & = & \sin\fr{3}{2}\Gt-\fr{1+\Gn}{\sqrt{2}(1+\Gn-\Gl^2)}\left(\sqrt{1-\lambda^2 \sin^2\Gt}+2\cos\Gt\right)\sqrt{\sqrt{1-\lambda^2 \sin^2\Gt}-\cos\Gt}.\label{H32}\nonumber
\end{eqnarray}

The proposed procedure, based on the assumption $1\le s<3$ and then on the analysis of the leading order term of the governing equation \eq{eq_lead}, can in principle be performed 
considering a different range of values of $s$, as $1 <  s \leq 3$, corresponding to terms $s=3/2,2,5/2,3$, but observing the expression \eq{Jsing} and remembering that the other
contributions to the J-integral are of orders $r^{s_i+s_j-1}$ and $r^{s_i+s_j-3}$ and that $s=1/2$ is excluded because finite energy is required, it is easy to deduce that all terms
associated to $s>5/2$ do not contribute to the energy release rate. Therefore we can conclude that $s=3/2$ provides effectively the only term contributing to the J-integral and that 
the choice of considering the leading term of the governing equation \eq{eq_steady} in the asymptotic analysis is correct for our purpose.

In order to evaluate the energy release rate \eq{J_32}, the constant $B_2$ must be explicitly determined. In the next Section $B_2$ is calculated starting from the asymptotic expansion of the full-field
solution derived in \citet{MishPicc1} for the same loading conditions \eq{bound+}-\eq{loading} by means of Wiener-Hopf technique.

\section{Explicit evaluation of the energy release rate}
\label{J_ev}

 In order to derive an explicit expression for the constant $B_2$, we perform the asymptotic analysis of the Fourier transform of the full-field solution derived
in \citet{MishPicc1} for the same loading conditions \eq{bound+}-\eq{loading} by means of Wiener-Hopf technique. In the limit $|s|\rightarrow \infty$, the Fourier transforms of stress, couple stress field and
displacements assume the following behavior:
\begin{eqnarray}
\ov{t}_{23}^+(s,0^+)  & = & \fr{FT_{0}\Xi(1+\eta-2h_0^2m^2)}{\Upsilon(h_0,m,\eta)}(s\ell)_+^{1/2}+O\left((s\ell)_+^{-1/2}\right),\quad \mbox{Im}s>0, \label{four1} \\
\ov{\mu}_{22}^+(s,0^+)& = & \fr{2iFT_{0}\ell\Xi\left(\sqrt{1-2h_0^2m^2}-\eta\right)(1+\eta)}{\Upsilon(h_0,m,\eta)\left(1+\sqrt{1-2h_0^2m^2}\right)}(s\ell)_+^{-1/2}+O\left((s\ell)_+^{-1}\right),\ \mbox{Im}s>0, \label{four12}\\
\ov{w}^-(s,0^+)       & = & -\fr{2FT_{0}\ell\Xi}{G\Upsilon(h_0,m,\eta)}(s\ell)_-^{-5/2}+O\left((s\ell)_-^{-7/2}\right),\quad \mbox{Im}s<0, \label{four2}
\end{eqnarray}
where $F$ is a constant determined starting from the the boundary condition \eq{bound+}$_{1}$ and applying the Liouville theorem and:
\beq
\Xi=\fr{k_{+}(i\ell/L)}{(i\ell/L)_{+}^{1/2}}, \quad \Upsilon(h_0,m,\eta)=\fr{1-\eta^2-2h_0^2m^2+2\sqrt{1-2h_0^2m^2}(1+\eta-h_0^2m^2)}{1+\sqrt{1-2h_0^2m^2}},
\eequ{factors}
the explicit expression for the factorization function $k_{+}(i\ell/L)$ is given by \citet{MishPicc1}. Substituting the \eq{factors}$_{(1)}$ into \eq{four1} and \eq{four2} we obtain:
\begin{eqnarray}
\ov{t}_{23}^+(s,0^+)  & = & \fr{FT_{0}k_{+}(i\ell/L)}{(i\ell/L)_{+}^{1/2}}\fr{(1+\eta-2h_0^2m^2)}{\Upsilon(h_0,m,\eta)}(s\ell)_+^{1/2}+O\left((s\ell)_+^{-1/2}\right),\quad \mbox{Im}s>0,\label{four11}\\
\ov{\mu}_{22}^+(s,0^+)& = & \fr{2iFT_{0}k_{+}(i\ell/L)\left(\sqrt{1-2h_0^2m^2}-\eta\right)(1+\eta)}{(i\ell/L)_{+}^{1/2}\Upsilon(h_0,m,\eta)\left(1+\sqrt{1-2h_0^2m^2}\right)}(s\ell)_+^{-1/2}+O\left((s\ell)_+^{-1}\right),\ \mbox{Im}s>0,\label{four12}\\
\ov{w}^-(s,0^+)       & = & -\fr{2FT_{0}\ell k_{+}(i\ell/L)}{G(i\ell/L)_{+}^{1/2}\Upsilon(h_0,m,\eta)}(s\ell)_-^{-5/2}+O\left((s\ell)_-^{-7/2}\right),\quad \mbox{Im}s<0.\label{four22}
\end{eqnarray}

Further, we consider the following transformation formula \citep{Roos1}:
\beq
 x^{\kappa}\stackrel{ft}{\leftrightarrow}i^{\kappa+1}\Gamma(\kappa+1)s^{-\kappa-1},\ \mbox{with}\ \kappa\neq-1,-2,-3\ldots,
\eequ{abel}
where $\Gamma$ is the gamma function and the symbol $\stackrel{ft}{\leftrightarrow}$ indicates that the quantities on the two sides of the \eq{abel}  are connected by means of unilateral Fourier transform.
Applying the \eq{abel} to expressions \eq{four11} and \eq{four22}, we get:
\begin{eqnarray}
t_{23}    (x,0^+)& = & -\fr{FT_{0}\sqrt{L}k_{+}(i\ell/L)}{2\sqrt{\pi}}\fr{(1+\eta-2h_0^2m^2)}{\Upsilon(h_0,m,\eta)}x^{-3/2},\quad x>0,\label{t23real}\\
\mu_{22}  (x,0^+)& = &   \fr{2FT_{0}\sqrt{L}k_{+}(i\ell/L)\left(\sqrt{1-2h_0^2m^2}-\eta\right)(1+\eta)}{\sqrt{\pi}\Upsilon(h_0,m,\eta)\left(1+\sqrt{1-2h_0^2m^2}\right)}x^{-1/2}, \ x>0,\label{mu22real}\\
w(x,0^+)         & = & \fr{8FT_{0}\sqrt{L} k_{+}(i\ell/L)}{3\sqrt{\pi}G\ell^2\Upsilon(h_0,m,\eta)}(-x)^{3/2},\quad x<0.\label{wreal}
\end{eqnarray}

On the crack surface, for $\theta=\pi$ and $y=0^+$, the term of order $3/2$ of the asymptotic expression in polar coordinates \eq{lead_w} becomes
\beq
w(x,0^+)=B_2 \fr{2 m^2 h_0^2}{1+\eta-2 m^2 h_0^2}(-x)^{3/2},\quad x<0,
\eequ{w0}
where the definition $r(y=0^+)=\sqrt{x^2}=|x|$ has been used. Equating expressions \eq{wreal} and \eq{w0}, we get:
\beq
B_2=\fr{4FT_{0}\sqrt{L}k_{+}(i\ell/L)}{3\sqrt{\pi}G\Upsilon(h_0,m,\eta)\ell^2}\left(\fr{1+\eta-2 m^2 h_0^2}{m^2 h_0^2}\right).
\eequ{B2}

The explicit expression \eq{B2} can the be used into the \eq{J_32} for studying the variation of the energy release rate in function of the crack tip speed and of the microstructural parameters
$h_0$ and $\eta$.

\begin{figure}[htbp]
\centering
\includegraphics[width=15cm,height=6.5cm]{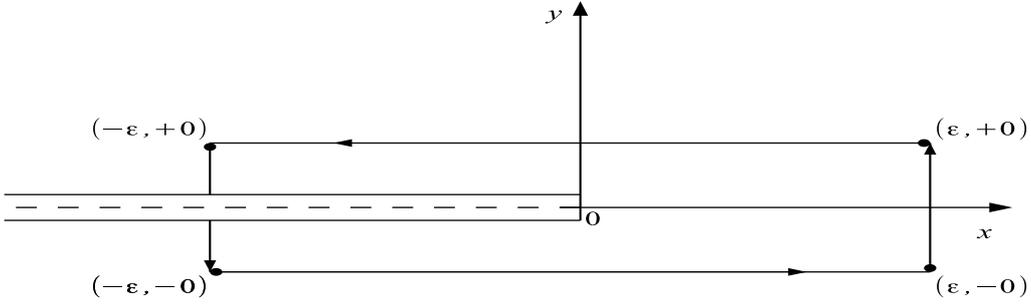}
\caption[square_path]{\footnotesize {Rectangular-shaped contour around the crack-tip}}
\label{square_path}
\end{figure}

In order to check the validity of the obtained results, an alternative expression for the energy release rate is derived considering the rectangular-shaped contour
$\Gamma$ with vanishing height along the y-direction and with $\varepsilon\rightarrow +0$ reported in Fig.\ref{square_path} \citep{Freund1}. The Cartesian components of the outward unit vector
normal to $\Gamma$ are $\Bn=(n_x,n_y,0)$. Considering the moving framework in Fig.\ref{square_path} with the origin at the crack tip, for the steady-state antiplane crack problem the generalized J-integral \eq{J_intA} becomes:
\begin{eqnarray}
\CJ & = & \int_{\Gamma}\left[(W+T)n_x-(t_{13}n_x+t_{23}n_y)w_{,x}-(\Gm_{11}n_{x}+\Gm_{21}n_y)\Gvf_{1,x}-(\Gm_{12}n_{x}+\Gm_{22}n_y)\Gvf_{2,x}\right]ds=\nonumber\\
    & = & \int_{\Gamma}\left[(W+T)-t_{13}w_{,x}-(\Gm_{11}\Gvf_{1,x}+\Gm_{12}\Gvf_{2,x})\right]dy-\int_{\Gamma}\left[t_{23}w_{,x}+(\Gm_{21}\Gvf_{1,x}+\Gm_{22}\Gvf_{2,x})\right]dx,\label{Jcart}
\end{eqnarray}
where $t_{13}$ and $t_{23}$ are components of the total non-symmetric Cauchy stress tensor, including both symmetric and skew-symmetric part.

In order to evaluate the energy release rate, we allow the height of the rectangular path reported in Fig.\ref{square_path} to vanish.
In this limit, the first integral of the \eq{Jcart} is zero and the following expression for the dynamic energy release rate is derived:
\beq
\CE =-2\lim_{\Gve\rightarrow+0}\left\{\int_{-\Gve}^{+\Gve}\left[t_{23}w_{,x}+(\Gm_{21}\Gvf_{1,x}+\Gm_{22}\Gvf_{2,x})\right]dx\right\}.
\eequ{ERR_square}

It is important to note that anti-symmetry conditions \eq{bound+} together with boundary conditions \eq{bound-} provide that the reduced traction $q_1=\mu_{21}$ is zero along the whole crack line $y=0$,
where $\Bn=(0,\pm1, 0)$. Consequently, the energy release rate \eq{ERR_square} becomes:
\begin{eqnarray}
\CE & = & -2\lim_{\Gve\rightarrow+0}\left\{\int_{-\Gve}^{+\Gve}\left[t_{23}(x,0^+)w_{,x}(x,0^+)+\Gm_{21}(x,0^+)\Gvf_{1,x}(x,0^+)+\Gm_{22}(x,0^+)\Gvf_{2,x}(x,0^+)\right]dx\right\}=\nonumber\\
    & = & -2\lim_{\Gve\rightarrow+0}\left\{\int_{-\Gve}^{+\Gve}\left[t_{23}(x,0^+)w_{,x}(x,0^+)+\Gm_{22}(x,0^+)\Gvf_{2,x}(x,0^+)\right]dx\right\}\nonumber\\
    & = & -2\lim_{\Gve\rightarrow+0}\left\{\int_{-\Gve}^{+\Gve}\left[t_{23}(x,0^+)w_{,x}(x,0^+)-\fr{1}{2}\Gm_{22}(x,0^+)w_{,xx}(x,0^+)\right]dx\right\}.\label{J_rect}
\end{eqnarray}

For evaluating the integral \eq{J_rect}, solely asymptotic expressions for the traction ahead of the crack tip $t_{23}$, the couple stress field $\mu_{22}$, and the displacement $w$ are required.
Then, by substituting equations \eq{t23real}, \eq{mu22real} and \eq{wreal} in the general formula \eq{J_rect}, we finally derive: 
\begin{eqnarray}
\CE & = & -2\lim_{\Gve\rightarrow+0}\left\{\fr{2F^2T_0^2Lk^{2}_{+}(i\ell/L)}{\pi G\ell^2\Upsilon^2(h_0,m,\eta)}\left[(1+\eta-2h_0^2m^2)\int_{-\Gve}^{+\Gve}x_{-}^{1/2}x_{+}^{-3/2}dx-\right.\right.\nonumber\\
    & - & \left. \left. \fr{\left(\sqrt{1-h_0^2m^2}-\eta\right)\left(1+\eta\right)}{\left(\sqrt{1-h_0^2m^2}+1\right)}\int_{-\Gve}^{+\Gve}x_{-}^{-1/2}x_{+}^{-1/2}dx \right]\right\}\label{J_distr}
\end{eqnarray}
where $x_{-}^{1/2}, x_{-}^{-1/2}$ and $x_{+}^{-3/2}, x_{+}^{-1/2}$ are distributions of the bisection type \citep{Arfk1}. For any real $\alpha$ with the exception of $\alpha=1,2,3,\dots$,
this particular type of distributions is defined as follows:
$$
x_{+}^{\alpha} = \left\{ \begin{array}{cc}
|x|^\alpha , & \mbox{for}~ x  > 0, \\
0 , & \mbox{for} ~ x < 0.
\end{array} \right.,\
x_{-}^{\alpha} = \left\{ \begin{array}{cc}
0 , & \mbox{for}~ x > 0, \\
|x|^\alpha, & \mbox{for} ~ x < 0.
\end{array} \right.
$$
Moreover, the product of distributions inside the integral in \eq{J_distr} is evaluated through the application of Fisher's theorem \citep{Fisch1}, that leads to the operational relation:
\beq
(x_{-})^\alpha(x_{+})^{-1-\alpha}=-\fr{\pi\delta(x)}{2\sin(\pi\alpha)},\ \mbox{with} \ \alpha\neq-1,-2,-3\dots,
\eequ{fisher}
where $\delta(x)$ is the Dirac delta distribution. Then, by using the relation \eq{fisher} into \eq{J_distr} and considering the fundamental property of the Dirac delta distribution
$\int_{-\Gve}^{+\Gve}\delta(x)dx=1$ \citep{Arfk1}, we finally get:
\beq
\CE=\fr{2F^2T_0^2Lk_{+}^{2}(i\ell/L)}{G\ell^2\Upsilon^2(h_0,m,\eta)}\left(\fr{2(1+\eta-h_0^2m^2)\sqrt{1-2h_0^2m^2}+(1-2h_0^2m^2-\eta^2)}{\sqrt{1-2h_0^2m^2}+1}\right)=\fr{2F^2T_0^2Lk_{+}^{2}(i\ell/L)}{G\ell^2\Upsilon(h_0,m,\eta)}.
\eequ{J_expl}

Using this alternative procedure, we have derived the explicit expression \eq{J_expl} for the energy release rate corresponding to a Mode III steady state propagating crack in a 
couple stress elastic material. Equation \eq{J_expl} can be compared with the energy release rate associate to an antiplane steady state crack in a classical elastic material, 
derived assuming the same loading configuration
\eq{loading}:
\beq
\CE^{cl}=\fr{T_0^2}{GL}\fr{1}{\sqrt{1-m^2}},
\eequ{J_classic}
the ratio between the two expressions \eq{J_expl} and \eq{J_classic} is given by:
\beq
\fr{\CE}{\CE^{cl}}=\fr{2F^2L^2k^{2}_{+}(i\ell/L)}{\ell^2\Upsilon(h_0,m,\eta)}\sqrt{1-m^2}.
\eequ{J_ratio}

The analytical expression \eq{J_expl} has been proved to be equivalent to \eq{J_32} by means of several numerical examples. Indeed,
both expressions \eq{J_32} and \eq{J_expl} provide the same results, which are reported in the next Section.

\section{Results and discussion}
\label{results}
 In this Section, the variation of the energy release rate is analyzed applying the classical Griffith criterion \citep{Willis2} in order to study the propagation stability.
 The results are compared with those detected using the $t^{\textrm{max}}$ criterion, adopted by \cite{MishPicc1}.

\begin{figure}[!htcb]
\centering
\includegraphics[width=16cm]{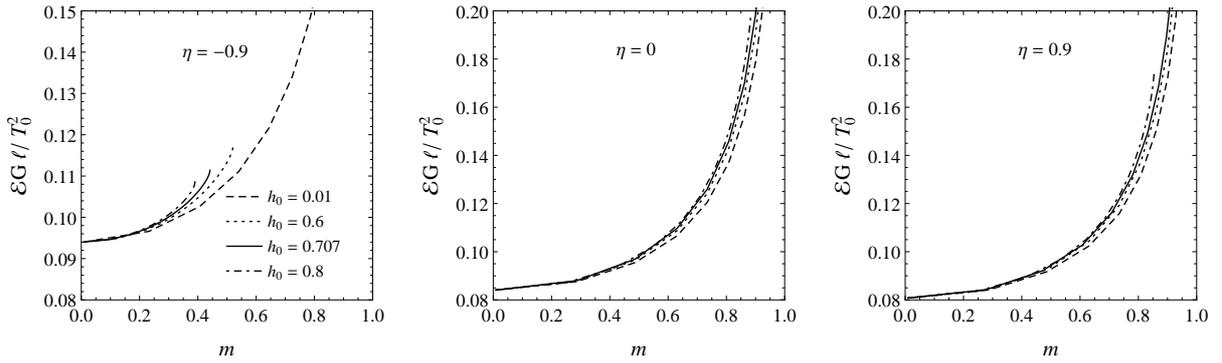}
\caption{\footnotesize Variation of the energy release rate as a function of the normalized crack tip speed $m$.}
\label{figjintA}
\end{figure}

 The normalized variation of the energy release rate versus the crack tip speed $m$ is reported in Fig. \ref{figjintA} for the same value of
 the ratio $L/\ell=10$, three different values of $\eta=\left\{0, \ 0.9, \ -0.9 \right\}$ and four different values of the rotational inertia $h_0=\left\{0.01,\ 0.6,\ 0.707,\ 0.8\right\}$. The range of
 the normalized crack tip velocity has been chosen in such a way that the propagation is subsonic and the conditions of validity of the full-field solution obtained in \cite{MishPicc1} 
 and used for evaluating the J-integral constant are satisfied. A similar behavior is observed for all different set of parameters: the energy release rate is initially constant for small values of the crack tip speed 
 $m \leq 0.3$, then increases monotonically until its limiting value. For small values of $h_0$ and $\eta$, the limit value of the energy release rate corresponds to $m=1$, 
 while as the microstructural parameters increases the limit value of the crack tip speed becomes smaller than the shear waves speed $c_s$, and thus $m<1$.\\
 
\begin{figure}[!htcb]
\centering
\includegraphics[width=16cm]{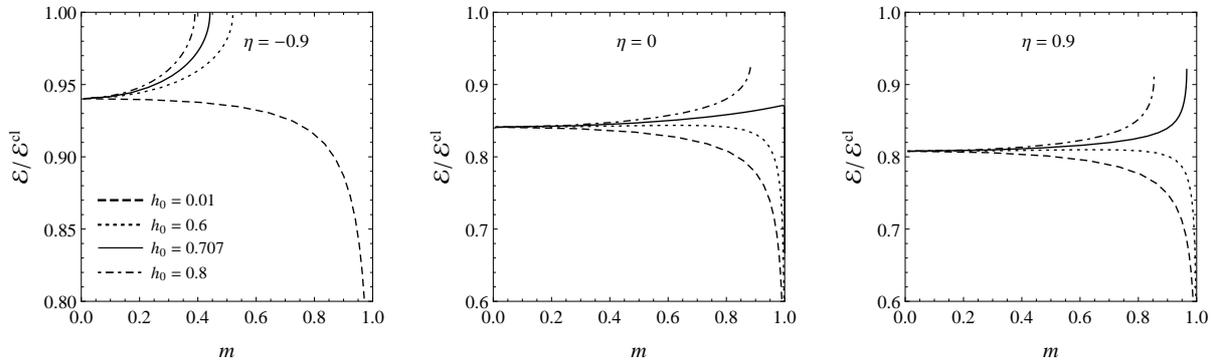}
\caption{\footnotesize Variation of the ratio $\CE/\CE^{cl}$ as a function of the normalized crack tip speed $m$.}
\label{figjintB}
\end{figure}
 
 On the basis of the Griffith criterion  \citep{ Freund1}, crack initiation requires that the energy release rate achieves a critical value $\CE_c=2\gamma$, where $\gamma$ is the energy needed to form a unit of new material
 surface, and is a constant depending only by the properties of the medium. In our case, once this critical value for the crack initiation is achieved, the energy release rate always increases in function
 of the velocity, this means that if the applied loading provides the energy necessary for starting the fracture process, the crack has enough energy to accelerate rapidly up to the limiting values
 of the speed \citep{Willis1, ObrMov1}. This implies that, if we analyze the results shown in Fig. \ref{figjintA} by means of Griffith criterion, the steady state propagation of a
 Mode III crack in couple-stress elastic material can be considered unstable for any value of the rotational inertia $h_0$ and of $\eta$. 
 
 The ratio between the energy release rate in couple stress materials and the energy release rate in classical elastic materials \eq{J_ratio} is plotted in Fig. \ref{figjintB} as a function of the normalized
 crack tip speed $m$. For small values of the rotational inertia, this ratio decreases as $m$ tends to $1$. This is due to the fact that for $m=1$, the energy release rate
 corresponding to a classical elastic material \eq{J_classic} diverges, while in presence of couple stress it has a finite limiting value. Differently, for large values
 of $h_0$, the ratio increases monotonically until a limiting value corresponding to $m<1$ and thus to a crack tip speed smaller than $c_s$. Therefore,
 in order to propagate the crack at constant velocity, more energy than in a classical elastic material must be provided by the loading if the contribution 
 of the rotational inertia is relevant. Observing Fig. \ref{figjintB}, we can also note that in the static limit,
 namely as $m$ tends to zero, the ratio $\CE/\CE^{cl}$ is close to one for $\eta=-1$, this behavior is in agreement with the results reported by \cite{Radi1}, which
 illustrate that for $\eta=-1$ the solution approaches the classical elastic case.
 
\begin{figure}[!htcb]
\centering
\includegraphics[width=8cm]{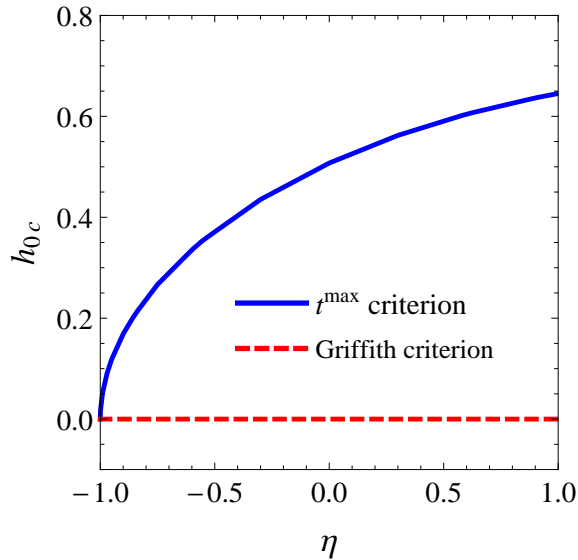}
\caption{\footnotesize Critical value of the rotational inertia $h_{0c}$ as a function of the ratio $\eta$.}
\label{h0critic}
\end{figure}

 In \cite{MishPicc1} the  $t^{\text{max}}$ criterion \citep{Geo1} is applied to the same crack problem. This alternative criterion states that the maximum total shear stress $t_{23}^{\text{max}}$ possesses a critical
 level $\tau_C$ at which the crack starts propagating. The behavior of  $t_{23}^{\text{max}}$ as a function of the crack tip speed has been studied extensively for several sets of microstructural parameters.
 For each values of the ratio $\eta$, it has been individuated a critical value of the rotational inertia $h_{0c}$, such that for $h_0 > h_{0c}$ the maximum shear stress increases very rapidly in function of $m$ 
 and becomes unbounded at a crack-tip speed lower than the shear wave speed $c_s$. In these cases the crack propagation turns out to be unstable and a limit speed of propagation $v_c<c_s$ is individuated. 
 For $h_0 < h_{0c}$, $t_{23}^{\text{max}}$ decreases instead as the crack speed increases and tends to zero as $m$ approaches its limiting value, suggesting the occurrence of stable crack propagation
 at velocities sufficiently lower than the shear wave speed. The critical value of the rotational inertia is reported as a function of $\eta$ in Fig.\ref{h0critic} (blue line).
 The limit value $\eta=-1$ corresponds to $h_{0c}=0$, then in this case the propagation turns out to be unstable for any value of the rotational inertia,
 while as $\eta$ increases $h_{0c}$ grows, and the range of the rotational inertia associated to stable cracks propagation becomes larger.  Therefore, the present analysis shows that
 if the maximum total shear stress criterion is adopted, a stabilizing effect of the crack propagation is provided as the characteristic ratio $\eta$ increases and then the 
 contribution of the microstructures effect becomes relevant. Differently, for negative values of $\eta$ near to the limit case $\eta=-1$, crack propagation is detected to 
 be unstable as the rotational inertia becomes not negligible. 
 
 The stabilizing effect is not detected applying the Griffith criterion, according to which the propagation is unstable regardless of the values of $\eta$ and $h_0$.
 As a consequence, assuming this criterion, the critical value of the rotational inertia is zero for all values of $\eta$ (dashed line
 in Fig. \ref{h0critic}), and the fracture is unstable for each set of microstructural parameters. This is due to the fact that the energy release rate is evaluated using the term of order $r^{3/2}$ 
 of the asymptotic displacement field, corresponding to the singular shear stress term of order $r^{-3/2}$. As it has been illustrated by many studies both in classical elastic materials 
 \citep{DuHanc1, DuBet1} and in couple stress \citep{RadiGei1, Radi1}, this singular contribution dominates very near to the crack tip, but it displays an unphysical negative shear stress ahead of the crack tip. Moreover,
 it is not sufficient to describe accurately the physical behavior of the stresses at a characteristic length from the crack tip where the higher order terms of the expansions 
 become important. In particular, the second term, which is associated to the linear term of the displacement and is commonly known as T-stress, can influence significantly the processes of crack 
 initiation and propagation in many physical situations \citep{TverHutch1}. In this cases, the critical stress intensity factor criterion and thus the connected Griffith criterion \citep{Freund1} are insufficient
 for describing accurately the crack extension and stability. Alternative two-parameters fracture criteria requiring the achievement of a critical value for both the stress 
 intensity factor and the T-stress has been proposed for classical elastic materials \citep{HancDu1,SmithAya1} and can be extended to couple stress.\\

 In couple stress materials, as the distance from the crack tip increases the discrepancy between the stresses physical behavior and the singular leading term of the asymptotic 
 is even more relevant respect to the classical elastic case \citep{Radi1}. Therefore, the contribution of the higher order term in the asymptotic shear stress, 
 which does not affect the energy release rate, is relevant. Differently, in \citet{MishPicc1} the total shear stress is
 evaluated by means of the full-field solution, that describes correctly the behavior of displacement and stresses on the whole crack line
 and takes fully into account the action of the microstructures.
 
 \begin{figure}[!htcb]
\centering
\includegraphics[width=14cm]{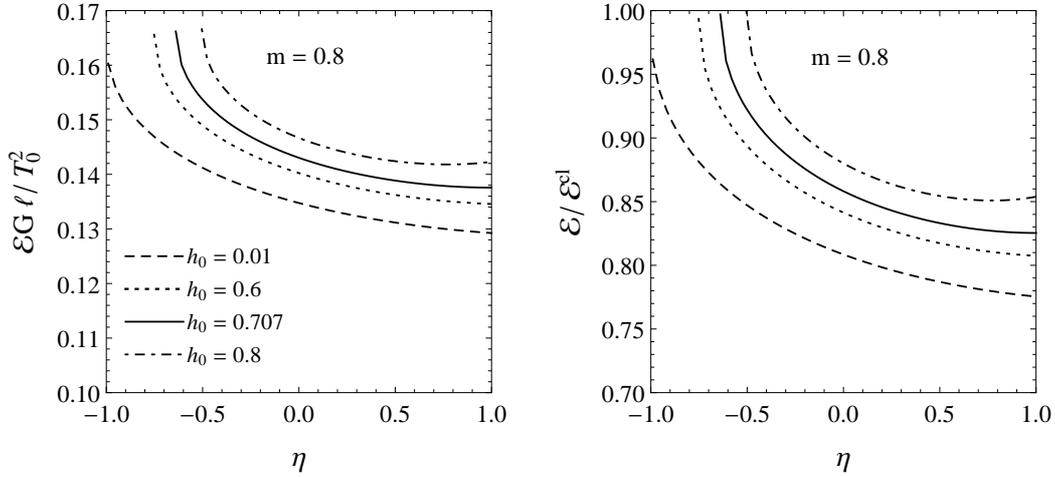}
\caption{\footnotesize Variation of the dynamic energy release rate and of the ratio $\CE/\CE^{cl}$ as a function of $\eta$ plotted for $m=0.8$.}
\label{figjintETA}
\end{figure}
\begin{figure}[!htcb]
\centering
\includegraphics[width=14cm]{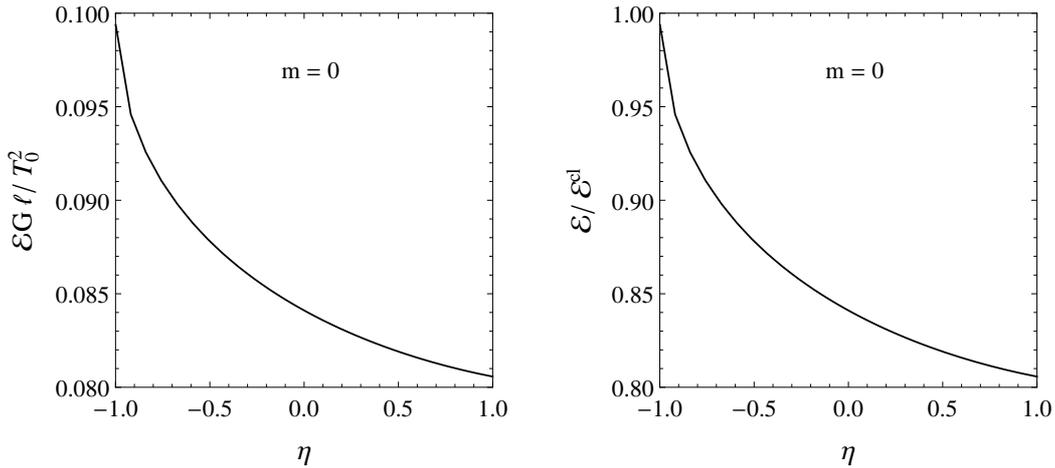}
\caption{\footnotesize Variation of the dynamic energy release rate and of the ratio $\CE/\CE^{cl}$ as a function of $\eta$ for the case $m=0$, corresponding to the static limit.}
\label{figjintETA0}
\end{figure}

 Fig. \ref{figjintETA} shows that for a fixed value of the crack tip speed, the energy release rate slightly increases its magnitude for large values of $h_0$. 
 As a consequence, for large values of the rotational inertia, a greater flux of energy at the crack tip is detected, and thus crack initiation and propagation are favored. Moreover, in  Fig. \ref{figjintETA} 
 we can also observe that ERR and the ratio $\CE/\CE_c$ decrease as $\eta$ increase. In particular, the ratio $\CE/\CE_c$ is almost one for $\eta=-1$, and then it decreases.
 As it is reported in Fig. \ref{figjintETA0}, this behavior is observed also in the static case, corresponding to the limit $m=0$. This confirms the results
 illustrated in \cite{Radi1} and can be explained with the fact that for large values of $\eta$, associated with large values of the characteristic length in torsion, the toughness 
 of the material increases, and thus less energy is provided at the crack tip and a shielding effect due to microstructures is detected. It is important to note that in the static case the energy release rate is 
 independent by the value of $h_0$, because the rotational inertia appears only in the dynamical terms of the evolution equation \eq{eq_steady} and of the J-integral \eq{Jex}.\\
 
\section{Conclusions}
A general expression for the J-integral associated to dynamic steady-state cracks subjected to antiplane loading in couple-stress elastic materials has been derived. The effects of both finite characteristics 
lengths as well as rotational inertia are included in the analysis. The generalized J-integral has been demonstrated to be path independent for steadily propagating cracks.
Asymptotic displacements and stress fields have been derived and used for evaluating the dynamic energy release rate. Surprisingly, the dependence of the energy release rate by the microstructural parameters 
looks less than the authors expectations. However, for large values of the characteristic length in torsion the ERR decreases, and thus less energy is provided at the crack tip and a shielding effect due to microstructures
is detected. 

The stability of the propagation has been studied by applying the energy-based Griffith criterion  \citep{Willis2, Willis1, ObrMov1}.
The propagation turns out to be unstable for each value of the characteristic length in torsion and of the rotational inertia. This result appears to contrast with those 
detected in \citet{MishPicc1}, where the maximum total shear stress criterion has been adopted, and a stabilizing effect in correspondence 
of relevant microstructural contributions is shown. In the authors' opinion, the discrepancy is due to the fact that the energy release rate depends only on a single term of the asymptotic 
displacement field, corresponding to the singular leading order term of the stresses. The singular stress dominates in proximity of the crack tip, whereas as the distance from the crack tip 
increases the effects of higher order asymptotic terms which do not contribute to the J-integral become relevant. It is important to note that
analogous results has been detected by comparing the ERR and the T-stress criterion in classical elasticity \citep{DuBet1,SmithAya1}. Moreover, the presence of the microstructures affects stresses and displacement 
up to a distance from the crack tip of the order of 5-10 times the characteristic length $\ell$ \citep{Radi1}, and thus cannot be described 
by the sole leading order term. Differently, the total shear stress is calculated by means of the full-field solution, that takes fully into account the microstructural contributions. Therefore, 
in order to study cracks propagation stability in elastic media with microstructures, fracture criteria considering also higher order terms must be developed and validated by means of experimental analysis. 
The $t^{\text{max}}$ criterion may represent one of the possible alternative methods that could give relevant results concerning microstructural effects on crack propagation stability.

\section*{Acknowledgements}
The authors are grateful to Dr. P. Gourgiotis for fruitful discussions and valuable suggestions. In particular, results reported in Appendix B should be attributed to him.
A. P. and L. M. gratefully thank financial support of the Italian Ministry of Education, University and Research in the framework of the 
FIRB project 2010 ``Structural mechanics models for renewable energy applications'', G. M. gratefully acknowledges the support from the European 
Union Seventh Framework Programme under contract number PIAP-GA-2011-286110-INTERCER2, and E. R. gratefully thanks financial support from the
``Cassa di Risparmio di Modena'' in the framework of the International Research Project $2009-2010$ ``Modelling of crack propagation in complex material''.

\newpage

\section*{Supplementary Material}
In this Section a general asymptotic solution for the evolution equation \eq{eq_lead} is derived. Equation \eq{eq_lead} can be split in the following two PDEs
\beq
\GD w=0, \qquad  \GD w-\Gl^2w_{,xx}=0.
\eequ{eq_split}
Substituting the asymptotic expression for the displacement \eq{asym} into \eq{eq_split}$_1$, and using the derivative rules \eq{deriv}, the following ODE for the unknown function $F_s(\Gt)$ is derived:
\beq
F_{s}^{''}(\Gt)+s^2F_s(\Gt)=0.
\eequ{eq_w1}
This equation admits the solutions providing:
\beq
w^I(r,\Gt)=r^s[C_1\cos(s\Gt)+C_2\sin(s\Gt)], 
\eequ{w1}
for any $s>0$.

Equation  \eq{eq_split}$_2$ can be reduced in the form \eq{eq_split}$_1$ by considering the linear coordinate transformation
\beq
X=x, \qquad Y=y\sqrt{1-\Gl^2}.
\eequ{coord_trasf}
Let us denote the transformed coordinates as $X=R\cos\GF$ and $Y=R\sin\GF$, where:
\beq
R(r,\Gt)=\sqrt{X^2+Y^2}=\sqrt{X^2+(1-\Gl^2)Y^2}=r\sqrt{1-\Gl^2\sin^2\Gt},
\eequ{R}
\beq
\GF(\Gt)=\arctan\left(\fr{Y}{X}\right)=\arcsin\left(\fr{\sqrt{1-\Gl^2}\sin\Gt}{\sqrt{1-\Gl^2\sin^2\Gt}}\right).
\eequ{Phi}
Then, the general solution of the equation \eq{eq_split}$_2$ is given by
\beq
w^{II}(r,\Gt)=R^s[C_3\cos(s\GF(\Gt))+C_4\sin(s\GF(\Gt))], 
\eequ{w2}
for any $s>0$.

Finally, the general solution of the governing equation \eq{eq_lead} is obtained by the sum of $w^{I}$ and $w^{II}$, given respectively by \eq{w1} and \eq{w2}. By using expression \eq{Phi}, we get:
\beq
w(r,\Gt)=r^s[C_1\cos(s\Gt)+C_2\sin(s\Gt)+(1-\Gl^2\sin^2\Gt)^{s/2}(C_3\cos(s\Phi(\Gt))+C_4\sin(s\Phi(\Gt)))].
\eequ{w_gen}

 Since the asymptotic expansion of the loading function \eq{loading} contains only integer powers of $r$, for non-integer values of the exponent $s$,
 which in our case means for $s \neq 1,2$,  the boundary conditions \eq{bound+} at $\theta=0$ and \eq{bound-} at $\theta=\pi$ yield the following linear and homogeneous system for the unknown
 constants $C_1, C_2, C_3$ and $C_4$, equivalent to that derived for the traction-free problem \citep{RadiGei1, Radi1}:
\begin{equation}
\label{system}
\begin{array}{l}
C_1+C_3=0, \\[3mm]
s(s-1) [(1 + \eta)C_1 + (1 + \eta - \lambda^2)C_3] = 0, \\[3mm]
s(s-1) \left\{ [(1 + \Gn)C_1 + (1 + \Gn - \Gl^2)C_3]\cos(s\pi) + [(1 + \Gn)C_2 + (1 + \Gn - \Gl^2)C_4]\sin(s\pi) \right\} = 0, \\[3mm]
\begin{array}{l}
s(s-1)(s-2) \left\{ [(1 + \Gn - \Gl^2)C_2 + \sqrt{1 - \Gl^2}(1 + \Gn)C_4]\cos(s\pi) \right. \\
\hspace{63mm} \left. - [(1 + \Gn - \Gl^2)C_1 + \sqrt{1 - \Gl^2}(1 + \Gn)C_3]\sin(s\pi) \right\} = 0.
\end{array}
\end{array}
\end{equation}
Then from \eqref{system} it necessarily follows that $C_1 = C_3 = 0$ and
\begin{equation}
\label{system2}
\begin{array}{l}
[(1 + \Gn)C_2 + (1 + \Gn - \Gl^2)C_4]\sin(s\pi) = 0, \\[3mm]
[(1 + \Gn - \Gl^2)C_2 + \sqrt{1 - \Gl^2}(1 + \Gn)C_4]\cos(s\pi) = 0.
\end{array}
\end{equation}
Eqs. \eqref{system2} admit a non trivial solution for $\sin 2\pi s = 0$, namely for $s = n/2$ where $n \in \mathbb{N}$. In particular,
for $n$ odd one obtains from \eqref{system2}$_1$:
\begin{equation}
C_4 = -\frac{1 + \eta}{1 + \eta - \lambda^2} C_2,
\end{equation}
and consequently from \eqref{w_gen}
\begin{equation}
w(r,\theta) = C_2 r^{n/2} \left[ \sin \left(\frac{n}{2}\theta\right) -
\frac{1 + \eta}{1 + \eta - \lambda^2} (1 - \lambda^2 \sin^2 \theta)^{n/4} \sin \left(\frac{n}{2}\Phi(\Gt)\right) \right].
\label{eq_odd}
\end{equation}
For $n$ even, from \eqref{system2}$_2$ it follows:
\begin{equation}
C_4 = -\frac{1 + \eta - \lambda^2}{\sqrt{1 - \lambda^2}(1 + \eta)} C_2,
\end{equation}
and consequently from \eqref{w_gen}
\begin{equation}
w(r,\theta) = C_2 r^{n/2} \left[ \sin \left(\frac{n}{2}\theta\right) -
\frac{1 + \eta - \lambda^2}{\sqrt{1 - \lambda^2}(1 + \eta)} (1 - \lambda^2 \sin^2 \theta)^{n/4} \sin \left(\frac{n}{2}\Phi(\Gt)\right) \right].
\end{equation}

Since we are considering values in the range $1\le s<3$, the terms corresponding to $s= 1, 3/2, 2, 5/2$ need to be included in the asymptotic solution. The terms of the order $s=3/2 $ and $s=5/2$ possess the
form (\ref{eq_odd}), while the terms $s=1, 2$ correspond to degenerate cases of the equation \eq{eq_lead}, and need to be treated separately.

For $s=1$, the general solution of the equation  \eq{eq_lead} may be found by solving the problem
\beq
\GD w=r^{-1} (1 - \lambda^2\sin^2\theta)^{-1/2} (C_3 \cos \Phi(\Gt) - C_4 \sin \Phi(\Gt)),
\eequ{s1}
namely
\beq
\GD w=r^{-1} (1 - \lambda^2\sin^2\theta)^{-1} (C_3 \cos \Gt - C4 \sqrt{1-\Gl^2}\sin \Gt),
\eequ{s1simple}

By using the method of variation of the parameters, one can find the following solution of equation \eq{s1simple} in the separate variable form \eqref{asym} 
\begin{eqnarray}
w(r,\theta) & = & \fr{r}{2\Gl^2}\left\{\left[C_1+C_3\log(1-\Gl^2\sin^2 \Gt)+2C_4(\GF-\Gt\sqrt{1-\Gl^2})\right]\cos\Gt+\right. \nonumber\\
            & + & \left.\left[C_2-2C_3(\GF\sqrt{1-\Gl^2}-\Gt)+C_4\sqrt{1-\Gl^2}\log(1-\Gl^2\sin^2\Gt)\right]\sin\Gt\right\}.\label{deg1}
\end{eqnarray}

The boundary conditions \eqref{bound+} and \eqref{bound-} necessarily imply that $C_1=C_3=C_4 = 0$, so that the corresponding solution reduces to
\begin{equation}
w(r,\theta) = C_2r \sin\theta,
\end{equation}
where $C_2$ depends by the amplitude of the loading \eq{loading} and also by the constant associated to the term $s=3$, that is neglected in our analysis.

For $s=2$, the general solution of the equation  \eq{eq_lead} may be found by solving the problem
\beq
\GD w=C_3+ C_4 \Phi(\Gt),
\eequ{s2}
namely
\beq
\GD w=C_3 + C_4 \arcsin\left(\fr{\sqrt{1-\Gl^2}\sin\Gt}{\sqrt{1-\Gl^2\sin^2\Gt}}\right),
\eequ{s2simple}

By using the method of variation of the parameters, one can find the following solution of equation \eq{s2simple} in the separate variable form \eqref{asym} 
\begin{eqnarray}
w(r,\theta)& = & r^2\left\{C_1\cos2\Gt+C_2\sin2\Gt+\fr{C_3}{4}+\right.\\
           & + & \left.\fr{C_4}{2\Gl^2}\left[(\Gl^2\sin^2\Gt+\cos2\Gt)\GF+\sqrt{1-\Gl^2}\left(\sin2\Gt\log\sqrt{2(1-\Gl^2\sin^2\Gt)}-\Gt\cos2\Gt\right)\right]\right\}.\nonumber
\end{eqnarray}

The boundary conditions \eqref{bound+} and \eqref{bound-} necessarily imply that $C_1=C_3=C_4 = 0$, so that the corresponding solution reduces to
\begin{equation}
w(r,\theta) = C_2r^2 \sin2\theta.
\end{equation}
where $C_2$ depends by the amplitude of the loading \eq{loading} and also by the constant associated to the term $s=4$, that is neglected in our analysis.
\section*{Appendix A}
In this Appendix, we demonstrate that the dynamic J-integral expression \eq{J_intA} for a Mode III steady-state propagating crack in couple stress elastic materials is
path independent.

Considering a closed oriented path formed by two crack tip contours $\Gamma_1$ and $\Gamma_2$ and by the segments of the crack faces of length $d$
that connect the ends of these contours, the energy flux integral corresponding to this entire closed path $\Gamma_{tot}$ for a Mode III steady state crack in
couple-stress materials is given by
\beq
F(\Gamma_{tot})=F(\Gamma_2)-F(\Gamma_1)=v\oint_{\Gamma_{tot}}\left[(W+T)n_{x}-\mathbf{t}^T\mathbf{n}\cdot\mathbf{e}_{z} w_{,x}-\BGm^T\mathbf{n}\cdot\BGvf_{,x}\right]ds,
\label{flux_3app}
\eeq
then from the definition \eq{J_intA} we derive
\beq
\mathcal{J}(\Gamma_{tot})=\mathcal{J}(\Gamma_2)-\mathcal{J}(\Gamma_1)
=\oint_{\Gamma_{tot}}\left[(W+T)n_{x}-\mathbf{t}^T\mathbf{n}\cdot\mathbf{e}_z w_{,x}-\BGm^T\mathbf{n}\cdot\BGvf_{,x}\right]ds,
\label{Jdelta}
\eeq
where the notation $\mathcal{J}(\Gamma_1)$ and $\mathcal{J}(\Gamma_2)$ denotes that the dynamic J-integral \eq{J_intA} is evaluated respect to the crack tip contours $\Gamma_1$ and $\Gamma_2$, respectively.
Applying the divergence theorem to the \eq{Jdelta} we obtain and remembering that $n_{x}=\mathbf{n}\cdot \mathbf{e}_{x}$, we obtain
\beq
\mathcal{J}\Gamma_2)-\mathcal{J}(\Gamma_1)=\int_{A_{tot}}\nabla\cdot\left[(W+T)\mathbf{e}_{x}-\mathbf{t}^T\mathbf{e}_z w_{,x}-\BGm^T\BGvf_{,x}\right]dA,
\label{Jdiv}
\eeq
where $A_{tot}$ is the area within the closed area. For the antiplane steady-state problem the strain elastic energy density and the kinetic energy density
are given by
\beq
W=\frac{1}{2}(\Gs_{13}w_{,x}+\Gs_{23}w_{,y}+\Gm_{11}\Gvf_{1,x}+\Gm_{12}\Gvf_{2,x}+\Gm_{21}\Gvf_{1,y}+\Gm_{22}\Gvf_{2,y}),
\label{Wen}
\eeq
\beq
T=\frac{v^2}{2}(\rho w_{,x}^2+J\Gvf_{1,x}^2+J\Gvf_{2,x}^2),
\label{Ken}
\eeq
the first term of the integral \eq{Jdiv} is the given by
\begin{eqnarray}
\nabla\cdot[(W+T)\mathbf{e}_x]   & = & (W+T)_{,x} = v^2\left(\rho w_{,xx}w_{,x}+J\Gvf_{1,xx}\Gvf_{1,x}+J\Gvf_{2,xx}\Gvf_{2,x}\right)+\nonumber\\
                                 & + & \frac{1}{2}\left(\Gs_{13,x}w_{,x}+\Gs_{13}w_{,xx}+\Gs_{23,x}w_{,y}+\Gs_{23}w_{3,yx}\right)+\nonumber\\
                                 & + & \frac{1}{2}\left(\Gm_{11,x}\Gvf_{1,x}+\Gm_{11}\Gvf_{1,xx}+\Gm_{12,x}\Gvf_{2,x}+\Gm_{12}\Gvf_{2,xx}+\right.\label{Jterm1}\\
                                 & + & \left.\Gm_{21,x}\Gvf_{1,y}+\Gm_{21}\Gvf_{1,yx}+\Gm_{22,x}\Gvf_{2,y}+\Gm_{22}\Gvf_{2,yx}\right).\nonumber
\end{eqnarray}
Taking into account the dynamic equilibrium conditions \eq{balance_eq}, the second term can be written as follows
\begin{eqnarray}
\nabla\cdot(\mathbf{t}^T\mathbf{e}_z w_{,x})& = & (\nabla\cdot\mathbf{t}^T)\cdot\mathbf{e}_z w_{,x}+\mathbf{t}^T\cdot\nabla w_{,x}=\nonumber\\
                                            & = & \rho\ddot{u}w_{,x}+\left(\Gs_{13}+\tau_{13}\right)w_{,xx}+\left(\Gs_{23}+\tau_{23}\right)w_{,yx}=\\ \label{Jterm2}
                                            & = & \rho v^2w_{,xx}w_{,x}+\left(\Gs_{13}+\tau_{13}\right)w_{,xx}+\left(\Gs_{23}+\tau_{23}\right)w_{,yx},\nonumber
\end{eqnarray}
while the third
\begin{eqnarray}
\nabla\cdot(\BGm^T\BGvf_{,x}) & = & (\nabla\cdot\BGm^T)\cdot\BGvf_{,x}+\BGm^T\cdot\nabla\BGvf_{,x}=\nonumber\\
                              & = &  \left(J\ddot{\Gvf}_1-2\tau_{23}\right)\Gvf_{1,x}+\left(J\ddot{\Gvf}_2+2\tau_{13}\right)\Gvf_{2,x}
                              +\Gm_{11}\Gvf_{1,x}+\Gm_{12}\Gvf_{2,x}+\Gm_{21}\Gvf_{1,y}+\Gm_{22}\Gvf_{2,y}=\label{Jterm3}\\
                              & = & \left(Jv^2\Gvf_{1,xx}-2\tau_{23}\right)\Gvf_{1,x}+\left(Jv^2\Gvf_{2,xx}+2\tau_{13}\right)\Gvf_{2,x}+
                              \Gm_{11}\Gvf_{1,x}+\Gm_{12}\Gvf_{2,x}+\Gm_{21}\Gvf_{1,y}+\Gm_{22}\Gvf_{2,y}.\nonumber
\end{eqnarray}
Substituting \eq{Jterm1}, \eq{Jterm2} and \eq{Jterm3} into the integral \eq{Jdiv} and writing $\Gvf_1$ and $\Gvf_2$ as functions of the displacement
by means of relations \eq{eps_phi}$_{(3,4)}$, we obtain
\begin{eqnarray}
\mathcal{J}(\Gamma_2)-\mathcal{J}(\Gamma_1) & = & \int_{A_{tot}}\left[\frac{1}{2}\left(\Gs_{13,x}w_{,x}+\Gs_{23,x}w_{,y}+\frac{1}{2}\left(\Gm_{11,x}w_{,yx}
-\Gm_{12,x}w_{,xx}+\Gm_{21,x}w_{,yy}-\Gm_{22,x}w_{,xy} \right)\right)-\right.\nonumber\\
                                            & - & \left.\frac{1}{2}\left(\Gs_{13}w_{,xx}+\Gs_{23}w_{,yx}+\frac{1}{2}\left(\Gm_{11}w_{,yxx}-\Gm_{12}w_{,xxx}
                                            +\Gm_{21}w_{,yyx}-\Gm_{22}w_{,xyx}\right)\right)\right]dA,\label{Jsim2}
\end{eqnarray}
finally, introducing into the \eq{Jsim2} expressions \eq{sigma} and \eq{mu}, defining the stress and couple-stress tensors in function of the derivatives of the displacement, we get:
\begin{eqnarray}
\mathcal{J}(\Gamma_2)-\mathcal{J}(\Gamma_1) & = & \frac{G\ell^2}{2}\int_{A_{tot}}\left[\left(w_{,xx}w_{,xxx}+w_{,yy}w_{,yyx}-\eta\left(w_{,yyx}w_{,xx}+w_{,xxx}w_{,yy}\right)\right)-\right.\nonumber\\
                                            & - & \left.\left(w_{,xx}w_{,xxx}+w_{,yy}w_{,yyx}-\eta\left(w_{,yyx}w_{,xx}+w_{,xxx}w_{,yy}\right)\right)\right]dA=0.
\end{eqnarray}

We have demonstrated that for a Mode III steady state crack propagation in couple-stress elastic materials the difference between the energy release rate calculated considering two different
paths around the crack tip is zero, as a consequence we can conclude that the J-integral expression \eq{J_intA} is path independent.
\section*{Appendix B}
In this Appendix we demonstrate that the two alternative expressions for the J-integral \eq{J_classic}, one of which is in function of the tractions and the other of the reduced tractions, are equivalent.
The equivalence of the two forms is demonstrated for a general three-dimensional steady state crack problem, and the expressions \eq{J_classic}, valid for the Mode III, are derived as a particular case.

On the basis of general expression for the energy flux \eq{CS_flux}, for a steady state three-dimensional crack propagating in couple stress elastic materials the J-integral is defined as follows:
\beq
\CJ = \int_{\Gamma}\left[(W+T)n_x-\Bt^T\Bn\cdot \Bu_{,x}+\BGm^T\Bn\cdot \BGvf_{,x}\right]ds.
\eequ{J_t}

An alternative expression for this integral is given in function of the reduced tractions $\Bp$ and of the couple stress tractions $\Bq$ \citep{Geo1, GourGeo1}:
\beq
\CJ =  \int_{\Gamma}\left[(W+T)n_x-\Bp \cdot \Bu_{,x}-\Bq\cdot \BGvf_{,x}\right]ds.
\eequ{J_p}
We now demonstrate that expressions \eq{J_t} and \eq{J_p} are equivalent. Remembering the definition of the reduced tractions and of the couple stress tractions \eq{trac}, the following relations are derived:
\begin{eqnarray}
 \Bp\cdot\Bu_{,x}   & = & \Bt^T\Bn\cdot \Bu_{,x}+\fr{1}{2}(\nabla\Gm_{nn}\times\Bn)\cdot \Bu_{,x} \nonumber\\
 \Bq\cdot\BGvf_{,x} & = & \BGm^T\Bn\cdot \BGvf_{,x}-\Gm_{nn}\Bn\cdot\BGvf_{,x}.
\end{eqnarray}

Making the difference between expressions \eq{J_t} and \eq{J_p}, we obtain:
\begin{displaymath}
 \fr{1}{2}\int_{\Gamma}\left[(\nabla\Gm_{nn}\times\Bn)\cdot \Bu_{,x}\right]ds-\int_{\Gamma}(\Gm_{nn}\Bn\cdot\BGvf_{,x})ds = \nonumber\\
\end{displaymath}
\beq
= \fr{1}{2}\Gve_{ijk}\int_{\Gamma}\left(\Gm_{nn,i}n_j u_{k,x}\right)ds-\fr{1}{2}\Gve_{kji}\int_{\Gamma}\left(\Gm_{nn}n_k u_{i,xj}\right)ds ,
\eequ{deltaJ1}
where $\Gve_{ijk}$ and $\Gve_{kji}$ are elements of the Levi-Civita tensor, and the components of the rotations vector $\BGvf$ have been expressed in function of the displacements according
to kinematic compatibility conditions introduced by \cite{Koit1} and reported in Section \ref{ss_crack} for the antiplane case. Taking into account the permutation properties of the Levi-Civita
tensor elements $\Gve_{kji}=\Gve_{jik}$ and $\Gve_{ijk}=-\Gve_{kji}$, the \eq{deltaJ1} becomes:
\begin{displaymath}
-\fr{1}{2}\Gve_{kji}\int_{\Gamma}\left(\Gm_{nn,i}n_j u_{k,x}\right)ds-\fr{1}{2}\Gve_{kji}\int_{\Gamma}\left(\Gm_{nn}n_k u_{i,xj}\right)ds =
\end{displaymath}
\begin{displaymath}
 =-\fr{1}{2}\Gve_{kji}\int_{\Gamma}\left(\Gm_{nn,i}n_j u_{k,x}\right)ds-\fr{1}{2}\Gve_{jik}\int_{\Gamma}\left(\Gm_{nn}n_j u_{k,xi}\right)ds=
\end{displaymath}
\beq
 =-\fr{1}{2}\Gve_{kji}\int_{\Gamma}\left(\Gm_{nn,i}n_j u_{k,x}+\Gm_{nn}n_j u_{k,xi}\right)ds=-\fr{1}{2}\int_{\Gamma}n_j\Gve_{kji}\left(\mu_{nn}u_{k,x}\right)_{,i}ds.
\eequ{deltaJ2}

Defining the vector $\Ba=\mu_{nn}\Bu_{,x}$ of components $a_k=\mu_{nn}u_{k,x}$, the \eq{deltaJ2} can be rewritten as:
\beq
 -\fr{1}{2}\int_{\Gamma}n_j\Gve_{kji}a_{k,i}ds=-\fr{1}{2}\int_{\Gamma}n_j\Gve_{jik}a_{k,i}ds=-\fr{1}{2}\int_{\Gamma}n_jb_{j}ds,
\eequ{deltaJ3}
where $b_j=\Gve_{jik}a_{k,i}$ are elements of the vector $\Bb=\mbox{rot}\Ba$. Applying the divergence theorem, it follows that:
\beq
-\fr{1}{2}\int_{\Gamma}\Bb\cdot\Bn ds=-\fr{1}{2}\int_{A}(\nabla\cdot\Bb) \ dA= -\fr{1}{2}\int_{A}[\nabla\cdot(\nabla\times \Ba)]\ dA=0.
\eequ{deltaJ3}

We have demonstrated that the difference between the two alternative forms of the J-integral \eq{J_t} and \eq{J_p} is zero, then expressions \eq{J_t} and \eq{J_p}
are equivalent. For a Mode III crack, using the definition of the out-of-plane displacement \eq{w}, the \eq{J_t} becomes:
\beq
\CJ= \int_{\Gamma}\left[(W+T)n_x-\Bt^T\Bn\cdot\Be_z w_{,x}-\BGm^T\Bn\cdot\BGvf_{,x}\right]ds.
\eequ{J_t3}
 In the same antiplane case, the following relations are derived:
\begin{eqnarray}
 \Bp\cdot\Bu_{,x}   & = & p_3 \cdot w_{,x}=(\Bp\cdot \Be_z) \nabla w\cdot \Be_x, \\
 \Bq\cdot \BGvf_{,x}& = & q_1\Gvf_{1,x}+q_2\Gvf_{2,x}=[(\nabla\BGvf)^T \Bq]\cdot \Be_x,
\end{eqnarray}
then, substituting these expressions in equation \eq{J_p}, we finally obtain the second form for the J-integral \eq{J_intB}:
\beq
\CJ= \int_{\Gamma}\left[(W+T)\Bn-\Bp \cdot\Be_z \nabla w-(\nabla \BGvf)^T \Bq\right]\cdot \Be_x ds
\eequ{J_p3}

\section*{Appendix C}

In this Appendix we derive the expression \eq{J_classic} for the energy release rate corresponding to a Mode III steady state propagating crack in a 
classical isotropic elastic material. For antiplane dynamical problems in classical elasticity the equation of motion \eq{eq_motion} becomes
\beq
G\GD u_3= \Gr\ddot{u}_{3}.
\eequ{eq_motion_cl}

Since we are interested in studying steady state crack propagation along $x_1-$axis, we perform the trasformation $u_3(x_1,x_2,t)=w(x,y)$ where $x=x_1-vt, y=x_2$,
(it is the same substitution illustrated by expression \eq{w}), and the  \eq{eq_motion_cl} then becomes:
\beq
(1-m^2)w_{,xx}+w_{,yy}=0,
\eequ{eq_motion_clw}
where $m=v/c_{s}$ and $c_{s}=\sqrt{G/\Gr}$. The Cauchy stresses are given by
\beq
\Gs_{13}=Gw_{,x}, \quad \Gs_{23}=Gw_{,y}.
\eequ{cauchy_cl}

The following conditions, equivalent to that imposed for couple stress materials (see expressions \eq{bound+} and  \eq{bound-}), are assumed on the crack surface, at $y=0$:
\begin{eqnarray}
\Gs_{23}(y=0)& = & -\tau(x), \quad -\infty<x<0, \label{bound_cl-}\\
w(y=0)       & = &  0,      \quad  0<x<+\infty, \label{bound_cl+}
\end{eqnarray}
where the same loading configuration considered for couple stress materials is applied in the crack faces:
\beq
\tau(x)=\fr{T_0}{L}e^{x/L}, \quad x<0.
\eequ{loading_cl}

An exact solution of the boundary value problem formulated will be obtained by means of Fourier transform and Wiener-Hopf technique. The direct and inverse Fourier transform of an arbitrary function $f(x)$
is defined as follows:
\beq
\ov{f}(s,y)=\int_{-\infty}^{+\infty}f(x,y)e^{isx}dx, \quad f(s,y)=\fr{1}{2\pi}\int_{L}\ov{f}(s,y)e^{-isx}ds,
\eequ{fourier_cl}
where $L$ denotes the inversion path within the region of analyticity of the function $\ov{f}(s,y)$ in the compx $s-$plane. Trasforming the evolution equation \eq{eq_motion_clw} we obtain the following ODE:
\beq
\ov{w}^{''}-s^2(1-m^2)\ov{w}=0,
\eequ{eq_clw_fou}
where the prime symbol $^{'}$ denotes the total derivative respect to $y$. The equation \eq{eq_clw_fou} possesses the following general solution that is required to be bounded as $y\rightarrow+\infty$:
\beq
\ov{w}(s,y)=B(s)e^{-\alpha(s)y},
\eequ{w_fou}
where $\alpha(s)=\sqrt{s^2(1-m^2)}$. The transformed stresses are given by:
\beq
\ov{\Gs}_{13}=-isG\ov{w}, \quad \ov{\Gs}_{23}=G\ov{w}^{'}.
\eequ{cauchy_cl_fou}

The Fourier transforms of the unknown stress ahead of the crack tip $\Gs_{23}(x>0,y=0)$ and of the crack faces displacements $w(x<0,y=0)$ are defined as follows:
\beq
\GS_{23}^+(s)= \int_{0}^{+\infty}\Gs_{23}(x,y=0)e^{isx}dx,
\eequ{Sigma_fou}
\beq
\Gs_{23}(x,y=0) = \fr{1}{2\pi}\int_{D}\GS_{23}^+(s)e^{-isx}ds, \quad x>0,
\eequ{Sigma_inv}
and
\beq
W^-(s)= \int_{-\infty}^{0}w(x,y=0)e^{isx}dx,
\eequ{W_fou}
\beq
w(x,y=0) = \fr{1}{2\pi}\int_{D}W^-(s)e^{-isx}ds, \quad x<0,
\eequ{W_inv}
where the inversion path is assumed to lie inside the region of analyticity of each transformed function. The transformed stress $\GS_{23}^+(s)$ is analytic and defined in the lower half complex
$s-$plane, $\mbox{Im}s<0$, whereas the transformed displacement $W^-(s)$ is analytic and defined in the upper half complex $s-$plane, $\mbox{Im}s>0$.

Taking into account the \eq{w_fou}, and substituting this expression into the \eq{cauchy_cl_fou}$_{(2)}$, in the limit $y\rightarrow 0$ we obtain:
\beq
B(s)=W^-(s), \quad \GS_{23}^+(s)=-\alpha(s)GW^-(s).
\eequ{SW}
As a consequence, equation \eq{SW} together with the condition \eq{bound_cl-} provides the following Wiener-Hopf equation connecting the two unkown functions $\GS_{23}^+(s)$ and  $W^-(s)$:
\beq
\GS_{23}^+(s)-\ov{\tau}^-(s)=-s_+^{1/2}s_-^{1/2}\nu GW^-(s),
\eequ{WH1}
where $\nu=\sqrt{1-m^2}$, $\ov{\tau}^-(s)$ is the Fourier transform of the loading function \eq{loading_cl}, defined in the lower half complex $s-$plane
\beq
\ov{\tau}^-(s)= \int_{-\infty}^{0}\tau(x)e^{isx}dx=\fr{T_0}{1+isL},
\eequ{tau_fou}
and the function $\sqrt{s^2}$ is factorized as follows \citep{MishPicc1}:
\beq
\sqrt{s^2}=s_+^{1/2}s_-^{1/2},
\eequ{s_pm}
where the functions $s_+$ and $s_-$ are analytic in the upper and in the lower half plane, respectively. Equation \eq{WH1} can then be rewritten as
\beq
\fr{\GS_{23}^+(s)}{s_+^{1/2}}=-s_-^{1/2}\nu GW^-(s)+\fr{T_0}{(1+isL)s_+^{1/2}}.
\eequ{WH2}

The second term in the right side of the Wiener-Hopf equation \eq{WH2} can be represented as
\beq
\Lambda(s)\equiv\fr{T_0}{(1+isL)s_+^{1/2}}=\Lambda^+(s)+\Lambda^-(s),
\eequ{WH_dec}
where
\beq
\Lambda^+(s)=\fr{T_0}{1+isL}\left[\fr{1}{s_+^{1/2}}-\fr{1}{(i/L)_+^{1/2}}\right],\quad \Lambda^-(s)=\fr{T_0}{1+isL}\left[\fr{1}{(i/L)_+^{1/2}}\right],
\eequ{Lambda}
and $\Lambda^+(s)$ is analytic function in the upper half plane, while $\Lambda^-(s)$ is an analytic function in the half plane $\mbox{Im}s<iL$. Using decomposition \eq{WH_dec}, the Wiener-Hopf equation
\eq{WH2} becomes
\beq
\fr{\GS_{23}^+(s)}{s_+^{1/2}}-\Lambda^+(s)=-s_-^{1/2}\nu GW^-(s)+\Lambda^-(s)\equiv E(s).
\eequ{WH_dec2}

The functional equation \eq{WH_dec2} defines the function $E(s)$ only in the real line. In order to evaluate this function, it is first necessary to examine the asymptotic behavior of the functions
$\GS_{23}^+(s)$ and $W^-(s)$. It has been demonstrated that for $x\rightarrow0\pm$ the stress and the displacement along the crack faces exhibit the following behavior:
\begin{eqnarray}
\Gs_{23}(x,y=0) & = &  O(x^{-1/2}) \ \mbox{as} \ x\rightarrow 0+,\label{asym1}\\
w(x,y=0)        & = &  O(x^{1/2})  \ \mbox{as} \ x\rightarrow 0-.\label{asym2}
\end{eqnarray}
Following the same procedure illustrated for couple stress materials, expressions \eq{asym1} and \eq{asym2} can be transformed by means of formula \eq{abel} applying Abel-Tauper type theorems \citep{Roos1}:
\begin{eqnarray}
\GS_{23}^+(s) & = &  O(s^{-1/2})  \ \mbox{as} \ |s|\rightarrow\infty \ \mbox{with} \ \mbox{Im}s>0  ,\label{asym1_tr}\\
W^-(s)        & = &  O(s^{-3/2})  \ \mbox{as} \ |s|\rightarrow\infty \ \mbox{with} \ \mbox{Im}s<0  .\label{asym2_tr}
\end{eqnarray}

Considering the asymptotic behavior of $\GS_{23}$ and $W^+$ and observing expressions \eq{Lambda}, we note that the first member of the Wiener-Hopf 
equation \eq{WH_dec2} is a bounded analytic function for $\mbox{Im}s>0$ that is zero as $|s|\rightarrow\infty$, whereas the second member is a bounded 
analytic function for $\mbox{Im}s<0$ that is also zero as $|s|\rightarrow\infty$. Then, for the theorem of analytic continuation, the two members define 
one and the same analytic function $E(s)$ over the entire complex $s-$plane. Moreover, Liouville's theorem leads to the conclusion
that $E(s)=0$. As a consequence, the transformed shear stress and displacement are given by:
\begin{eqnarray}
\GS_{23}^+(s) & = &  \Lambda^+(s)s_+^{1/2}, \ \mbox{Im}s>0 ,\label{WH_sigma}\\
W^-(s)        & = &  \fr{\Lambda^-(s)}{\nu Gs_-^{1/2}}, \  \mbox{Im}s<0.    \label{WH_w}
\end{eqnarray}
Evaluating the asymptotic leading term $|s|\rightarrow\infty$ of these expressions, we get:
\begin{eqnarray}
\GS_{23}^+(s) & = &  T_0(i/L)_+^{1/2}s_-^{-1/2}+O(s^{-1})  \ \mbox{as} \ |s|\rightarrow\infty\ \mbox{with} \ \mbox{Im}s>0  ,\label{asym1_sol}\\
W^-(s)        & = &  -\fr{T_{0}(i/L)_+^{1/2}}{\nu G}s_-^{-3/2} + O(s^{-2})\ \mbox{as} \ |s|\rightarrow\infty \ \mbox{with} \ \mbox{Im}s<0  ,\label{asym2_sol}
\end{eqnarray}
applying the transformation formula \eq{abel} to the \eq{asym1_sol} and \eq{asym2_sol} we finally obtain:
\begin{eqnarray}
\Gs_{23}(x,y=0) & = &  \fr{T_0}{\sqrt{\pi L}}x^{-1/2} \ \mbox{as} \ x\rightarrow 0+,\label{sigma_final}\\
w(x,y=0)        & = &  \fr{2T_0}{\nu G\sqrt{\pi L}}(-x)^{1/2}  \ \mbox{as} \ x\rightarrow 0-.\label{w_final}
\end{eqnarray}

The shear traction expression \eq{sigma_final} can then be used for calculating the stress intensity factor:
\beq
K_{III}^{cl}=\lim_{x \rightarrow 0}\sqrt{2\pi x}\Gs_{23}(x,y=0)=\sqrt{\fr{2}{L}}T_0.
\eequ{}

The dynamic J-integral for an antiplane steady state propagating crack is evaluated using the \eq{sigma_final} and \eq{w_final} and performing 
the same procedure illustrated for couple stress materials, choosing a rectangular shaped path surrounding the tip and applying the Fisher theorem:
\beq
\CE^{cl}=\fr{T_0^2}{\nu GL}=\fr{T_0^2}{GL}\fr{1}{\sqrt{1-m^2}}.
\eequ{J_clapp}

\newpage
\bibliography{WF_Bib} 
\bibliographystyle{elsarticle-harv}

\end{document}

%% file: macro.tex
\newcommand\eq[1] {(\ref{#1})}

\newcommand{\bfm}[1]{\mbox{\boldmath ${#1}$}}

\newcommand{\beqa}{\begin{eqnarray}}
\newcommand{\eeqa}[1]{\label{#1}\end{eqnarray}}
\newcommand{\bequ}{\begin{equation}}
\newcommand{\eequ}[1]{\label{#1}\end{equation}}

\newcommand{\ov}[1]{\overline{#1}}

\newcommand{\Gve}{\varepsilon}
\newcommand{\Gf}{\phi}
\newcommand{\Gvf}{\varphi}

\newcommand{\Gc}{\chi}

\newcommand{\Gl}{\lambda}
\newcommand{\Gn}{\eta}
\newcommand{\Gm}{\mu}

\newcommand{\Gt}{\theta}

\newcommand{\Gr}{\rho}

\newcommand{\Gs}{\sigma}

\newcommand{\Gj}{\tau}

\newcommand{\GD}{\Delta}
\newcommand{\GF}{\Phi}

\newcommand{\GS}{\Sigma}

\newcommand{\BGe}{\bfm\epsilon}
\newcommand{\BGve}{\bfm\varepsilon}

\newcommand{\BGvf}{\bfm\varphi}

\newcommand{\BGc}{\bfm\chi}

\newcommand{\BGm}{\bfm\mu}

\newcommand{\BGr}{\bfm\rho}

\newcommand{\BGs}{\bfm\sigma}

\newcommand{\BGj}{\bfm\tau}

\newcommand{\CE}{{\cal E}}

\newcommand{\CJ}{{\cal J}}

\def\Ba{{\bf a}}
\def\Bb{{\bf b}}

\def\Be{{\bf e}}

\def\Bn{{\bf n}}

\def\Bp{{\bf p}}
\def\Bq{{\bf q}}

\def\Bt{{\bf t}}
\def\Bu{{\bf u}}

\def\BI{{\bf I}}

\newcommand{\beq}{\begin{equation}}
\newcommand{\eeq}{\end{equation}}
\newcommand{\overliner}{\begin{eqnarray}}
\newcommand{\earr}{\end{eqnarray}}
\newcommand{\beqn}{\begin{equation*}}
\newcommand{\eeqn}{\end{equation*}}
\newcommand{\overlinern}{\begin{eqnarray*}}
\newcommand{\earrn}{\end{eqnarray*}}

\newcommand{\fr}{\frac}